\documentclass[nofootinbib,10pt,a4paper,twocolumn,english,tightenlines, showkeys]{revtex4}
\usepackage[T1]{fontenc}
\usepackage[latin9]{inputenc}
\usepackage{color}
\usepackage{textcomp}
\usepackage{amsmath}
\usepackage{amssymb}
\usepackage{graphicx}
\usepackage{epstopdf}
\usepackage{epsfig}
\date{}

\makeatletter

\providecommand{\tabularnewline}{\\}

\@ifundefined{date}{}{\date{}}
\usepackage{times}

\makeatother

\usepackage{babel}
\begin{document}

\title{Lower- and higher-order nonclassicality in a Bose-condensed optomechanical-like system and a Fabry\textendash Perot cavity with
one movable mirror: squeezing, antibunching and entanglement }

\author{Nasir Alam{$^{\dagger,}$}, Kishore Thapliyal{$^{\dagger}$},
Anirban Pathak{$^{\dagger,}$}\footnote[1]{anirban.pathak@gmail.com}, Biswajit Sen$^{\ddagger}$,
Amit Verma{$^{\star}$}, Swapan Mandal{$^{\mathsection}$}}

\affiliation{$^{\dagger}$Jaypee Institute of Information Technology, A-10, Sector-62,
Noida UP-201307, India~\\
$^{\star}$Jaypee Institute of Information Technology, Sector-128,
Noida, UP-201304, India~\\
$^{\ddagger}$Department of Physics, Vidyasagar Teachers' Training
College, Midnapore-721101, India~\\
$^{\mathsection}$Department of Physics, Visva-Bharati, Santiniketan-731235,
India}
\begin{abstract}
Various lower- and higher-order nonclassical properties have been
studied for two physical systems- (i) an optomechanical system composed
of a Fabry\textendash Perot cavity with one nonlinearly oscillating
mirror and (ii) an optomechanical-like system formed using a Bose-Einstein
condensate (BEC) trapped inside an optical cavity. The investigation
is performed using a perturbation method that leads to closed form
analytic expressions for the time evolution of the relevant bosonic
operators. In the first system, it is observed that the radiation
pressure coupling leads to the emergence of  lower- and higher-order
squeezing, antibunching, entanglement and intermodal squeezing. The
effects of the coherent interaction of a nonlinear oscillating mirror
with the cavity mode are studied, and it is observed that the optomechanical
system studied here becomes more nonclassical (entangled) when the
coupling strength is increased. It is also observed that the possibility
of observing entanglement depends on the phase of the movable mirror.
The Hamiltonian of the trapped BEC system is obtained as a special
case of the Hamiltonian of the first system, and the existence of
various nonclassicality in the trapped BEC system has been established,
and variations of those with various physical parameters have been
reported with an aim to understand the underlying physical process
that leads to and controls the nonclassicality.
\end{abstract}

\maketitle

\section{introduction}

Recent success in detecting gravitational wave \cite{ligo1,ligo2}
at the Laser Interferometer Gravitational-Wave Observatory (LIGO),
has enhanced the interest in the physical systems that can be used
to detect gravitational waves. A particularly important example of
such a physical system is an optomechanical system formed by movable
mirror of Fabry\textendash Perot cavity pumped by detuned laser \cite{squeezing-opto}.
In fact, the field of cavity optomechanics was originated from the
Braginsky and Khalili's pioneering proposal \cite{Gravitational_wave}
for the gravitational wave detection. Later on, several applications
have been reported for cavity optomechanics and related fields, and
various optomechanical \cite{BEC_SCience,BEC_Nature,key-3,key-9,S_Bose,key-4-1,opt-like3,opi-like4},
nanomechanical \cite{key-3,key-3-1,key-10} and optomechanics-like
\cite{BEC_Prl,opt-like1,opt-like2,opt-like5,opt-like6} systems have
been investigated both theoretically \cite{key-3,key-9,S_Bose,key-4-1,opt-like3}
and experimentally \cite{BEC_SCience,BEC_Nature,opi-like4,opt-like2}.
Interestingly, a majority of the recent investigations on the optomechanical
systems are focused on the nonclassical properties of the modes of
the optomechanical systems \cite{BEC_SCience,BEC_Nature,key-9,S_Bose,key-4-1,BEC_Prl}.
To stress on the relevance of these studies, it would be apt to note
that a nonclassical state does not have a classical analogue, and
it is characterized by the negative values of the Glauber-Sudarshan
$P$ function \cite{glaub,sudar}. The existence of nonclassicality
in general, and entanglement in particular has already been reported
in a nano-mechanical oscillator with a Cooper-pair box \cite{cooper pair},
ultracold atomic Bose-Einstein condensate (BEC) \cite{BEC,BEC 2},
arrays of nano-mechanical oscillators \cite{arrays nano osci}, two
mirrors of an optical ring cavity \cite{ring cavty}, or two mirrors
of two different cavities illuminated with entangled light beams \cite{2mr 2 cavity},
etc. The presence of other types of nonclassical states, like squeezed
and antibunched states has also been reported in optomechanical systems
\cite{key-3,antibunching-sps,OpmechSqueezing-1,OpmechSqueezing-2,OpmechSqueezing-3,OptomechAntibunch-1,OptomechAntibunch-2}.
Specifically, squeezed states are reported in Refs. \cite{key-3,OpmechSqueezing-1,OpmechSqueezing-2,OpmechSqueezing-3},
and antibunched states are reported in Refs. \cite{antibunching-sps,OptomechAntibunch-1,OptomechAntibunch-2}.
In many of these studies, Fabry\textendash Perot cavity played a crucial
role \cite{key-3,OpmechSqueezing-1,OpmechSqueezing-2,OptomechAntibunch-1,OptomechAntibunch-2},
and that is what motivated us for the present study.

Typically, it is assumed that one of the mirrors constituting the
Fabry\textendash Perot cavity oscillates linearly. However, Joshi
et al., have recently investigated a more general model that allows
nonlinear oscillation of the mirror \cite{key-3}. The model adopted
in \cite{key-3} appears to be more general, as the previously studied
Fabry\textendash Perot cavity having linearly oscillating mirror can
be obtained as a special case of it. Further, in the same limit (i.e.,
when the nonlinear part vanishes), the general Hamiltonian of the
Fabry\textendash Perot cavity with nonlinearly oscillating mirror
reduces to the Hamiltonian of an optomechanics-like system comprising
of a BEC trapped in an optical cavity, which has been recently studied
by various groups \cite{BEC_SCience,BEC_Nature,BEC_Prl}. In addition,
Mancini et al., \cite{key-4-1} have shown that the Schrodinger cat
states \cite{cat,cat 2} of the cavity field can be generated in a
Fabry\textendash Perot cavity with a movable mirror that can be treated
as quantum harmonic oscillators. The model of such a Fabry\textendash Perot
cavity using two two-level systems in a one dimensional waveguide
was proposed by Fratini et al., in Ref. \cite{tls}. The opto-mechanical
coupling, provided by radiation pressure, in various optomechanical
systems has been established as a useful tool for quantum state engineering
\cite{QuantumStateEngineering,QuantumStateEngineering-1,scissors}
as it can be used to manipulate the quantum state of light \cite{key-15,key-17,key-18}. 

The nonclassical states (i.e., quantum states having negative Glauber-Sudarshan
$P$ function) have various applications in quantum information processing
and other domains. Specifically, squeezed states have been used for
continuous variable quantum cryptography \cite{CV-qkd-hillery}, teleportation
of coherent states \cite{teleportation-of-coherent-state}, reduction
of noise in LIGO experiment \cite{sq-vac1,sq-vac2,sq-vac3}, etc.;
antibunched states are an essential ingredient of quantum cryptography
as it is useful in characterizing single photon sources \cite{antibunching-sps,IJP-anirban};
also, entangled states are essential for quantum teleportation \cite{Bennet1993},
densecoding \cite{densecoding}, entangled state-based quantum cryptography
\cite{Ekert-protocol}, etc. Due to their variety of applications,
characterization of nonclassical states is considered as an important
task. However, the Glauber-Sudarshan $P$ function is not directly
measurable through any experiment. Therefore, a set of moment based
criteria for nonclassical states have been introduced. Although only
an infinite set of these moment based criteria is both necessary and
sufficient as the $P$ function is, here we restrict our task to use
some of these criteria to establish the nonclassical behavior of the
optomechanical and optomechanics-like systems under consideration.
As the criteria used here are only sufficient (not necessary), satisfaction
of them would ensure the presence of the corresponding nonclassical
state, but the failure would not lead to any conclusion. In the analogy
of the lower-order criteria of nonclassicality, there exists some
criteria based on moments, which are functions of higher powers of
annihilation and creation operators and work as witnesses for higher-order
nonclassicality. Some recent experimental studies \cite{Maria-PRA-1,Maria-2,higher-order-PRL,with-Hamar}
have also established that sometimes it becomes easier to detect weak
nonclassicality using higher-order nonclassicality criteria in comparison
to their lower-order counterparts. Therefore, here we aim to study
both lower- and higher-order nonclassicality. 

The importance of Fabry\textendash Perot cavity and the potential
applications of nonclassical states discussed above have motivated
us to investigate the possibility of observing lower- and higher-order
nonclassicality in a Fabry\textendash Perot cavity with specific attention
to entanglement. In fact, this investigation is further motivated
by the recent works \cite{Gisin,Stannigel12} that clearly established
the relevance of cavity optomechanics in the interdisciplinary field
of quantum information processing. Specifically, in \cite{Stannigel12},
it was shown that preparation, storage and readout of heralded single
phonon Fock state is possible in a cavity optomechanical system, and
in \cite{Gisin}, it was established that the nonlinearities induced
in an optomechanical system (on or near resonance) can be used to
realize controlled quantum gates involving optical and phononic qubits.
This observation has specially motivated us to investigate the possibility
of observing intermodal entanglement in optomechanical systems.

\begin{figure}[h]
\begin{centering}
\includegraphics[angle=270,scale=0.45]{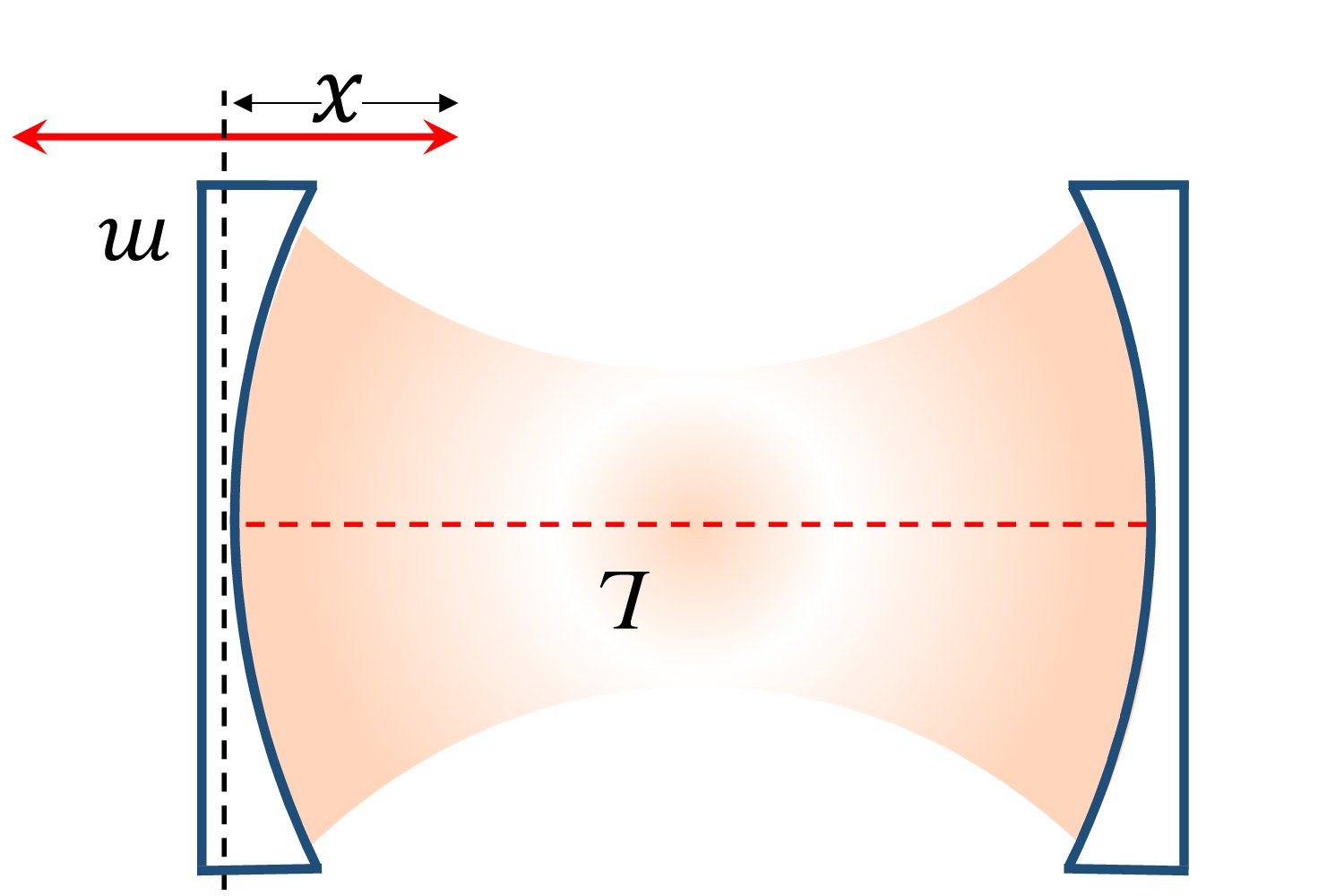}
\par\end{centering}
\caption{\label{fig:Fabry-Perot-cavity}Fabry\textendash Perot cavity with
one movable mirror having mass $m$, equilibrium cavity length $L$
and a maximum amplitude of vibrating mirror $x$.}
\end{figure}

Following independent approaches, nonclassical properties of BECs
are also studied in detail (\cite{BEC_SCience,BEC_Nature,BEC_Prl,AP_BEC-1,AP-BEC-2}
and references therein). Interestingly, in some of these studies,
efforts have been made to investigate optomechanics-like properties
of BECs, too \cite{BEC_SCience,BEC_Nature,BEC_Prl}. Specifically,
in Ref. \cite{BEC_Prl} investigations have been performed for a BEC
trapped inside an optical resonator and driven by both a classical
and a quantized light field; and very interestingly, a role reversal
between the matter-wave field and the quantized light field has been
observed. This role reversal phenomenon was of particular interest
as in this system, the matter-wave field (quantized light field) was
observed to play the role of the quantized light field (movable mirror),
and it was in sharp contrast to the earlier studied BEC-based cavity
optomechanical systems \cite{BEC_SCience,BEC_Nature}. This interesting
feature motivated us to investigate the nonclassical properties of
different modes of the BEC optomechanical system described in Ref.
\cite{BEC_Prl}, too. Specifically, in Ref. \cite{BEC_Prl}, the authors
studied the nonclassical nature of the system using Wigner function.
In the present study, we aim to study a more general system and obtain
the optomechanics-like system studied in \cite{BEC_Prl} as a special
case, and subsequently extend the results of \cite{BEC_Prl} by showing
the existence of lower- and higher-order nonclassicality via other
witnesses of nonclassicality.
In what follows, we investigate nonclassical properties of a system
composed of a Fabry\textendash Perot cavity with one movable mirror
\cite{key-3} as depicted in Fig. \ref{fig:Fabry-Perot-cavity}. At
first, we would consider the most general case in which the movable
mirror is nonlinear in nature with a nonlinearity proportional to
the $x^{4}$, where $x$ represents the displacement of the mirror
from its equilibrium position. The mirror is coupled to the field
mode of the cavity through the radiation pressure. Subsequently, we
would consider a special case of this system, where  the nonlinear
coupling constant vanishes and the system reduces to a system which
is mathematically equivalent to the BEC trapped optomechanics-like
system studied in Ref. \cite{BEC_Prl}. The coherent interaction of
the movable mirror with the cavity mode or trapped BEC is responsible
for lower- and higher-order nonclassical properties in the cavity
resonator. Such an optomechanical system is useful in various interdisciplinary
fields in quantum technology and quantum information \cite{tls}.
Consequently, numerous authors have previously treated the system
of a cavity field and a movable mirror, but no one has yet investigated
the possibilities of observing higher-order nonclassicalities. In
this article, we present the lower- and higher-order nonclassical
properties of the above mentioned optomechanical systems.

The rest of the paper is organized as follows. In Section \ref{sec2:the-model-hamiltonian},
we introduce the theoretical model describing the interaction between
a movable mirror with the quantized cavity mode in a Fabry\textendash Perot
cavity. We also describe how Hamiltonian of this system reduces to
that of the BEC trapped optomechanics-like system. In Section \ref{sec3:the-solution},
we provide a closed form analytic solution of the model Hamiltonian
in Heisenberg picture using a perturbative technique known as Sen-Mandal
technique. In Section \ref{sec4:squeezing}, we investigate the presence
of lower- and higher-order squeezing, and intermodal squeezing using
the field and oscillating mirror operators obtained in the previous
section. Similarly, in Section \ref{sec5:Antibunching}, we investigate
the possibility of observing lower- and higher-order antibunching
phenomena in the optomechanical system studied here. In Section \ref{sec6:entaglement},
we report the existence of lower- and higher- order entanglement using
various criteria. Finally, the paper is concluded in Section \ref{sec7:conclusion}.

\section{the model hamiltonian\label{sec2:the-model-hamiltonian}}

The optomechanical system of our interest is composed of a Fabry\textendash Perot
cavity with one fixed and one movable mirror. As stated in the previous
section, the movable mirror is nonlinear in nature with a nonlinearity
proportional to the $x^{4}$, where $x$ represents the displacement
of the mirror from its equilibrium position. This system models (imitates) a Kerr-like nonlinear medium illuminated with a
coherent light. It's possible to construct a quantum mechanical Hamiltonian
representing this system in a closed analytic form. In order to construct
such a Hamiltonian, the retardation effects due to the oscillating
mirror in the intracavity field are usually neglected. The correction
to the radiation pressure force due to the Doppler frequency shift
of the photons is also neglected \cite{doppler}. Further, the Casimir
effect \cite{casimir}, in the cavity, can also be safely neglected.
If we assume that the leakage of photons from the cavity is negligible,
then the main source of decoherence would be the coupling of mirrors
to its surroundings, which can also be neglected up to some extent
\cite{Singh}. Thus, neglecting the dissipation of the system, we
restrict ourselves to a situation where we only consider the unitary
time evolution of the coupled system of cavity and nonlinearly oscillating
mirror.

For further investigation, we express the model Hamiltonian in terms
of the creation and annihilation operators. On the application of
rotating wave approximation (RWA), i.e., neglecting the fast rotating
terms, the analytic form of the quantum mechanical Hamiltonian (see
Eq. 19 of \cite{key-3}) of the system described above takes the following
form 

\begin{equation}
\begin{array}{lcl}
H & = & \hbar\omega_{k}k^{\dagger}k+\hbar\omega_{m}a^{\dagger}a+\hbar\beta a^{\dagger2}a^{2}-\hbar g(k^{\dagger}ka^{\dagger}+{\rm H.c.}),\end{array}\label{eq:1}
\end{equation}
where $\omega_{m}=\omega_{0}+2\beta$ and H.c. stands for the Hermitian
conjugate. For simplicity, throughout the article we have used $\hbar=1$.
The annihilation (creation) operator $a\,(a^{\dagger})$ corresponds
to the mode of the movable mirror with frequency $\omega_{0}$ and
mass $m$ and the annihilation (creation) operator $k\,(k^{\dagger})$
corresponds to the quantized cavity mode with resonant frequency $\omega_{k}=n\frac{\pi c}{L}$,
where $n$ is an integer, $L$ is the equilibrium length of the cavity,
and $c$ is the velocity of light in the free space. The parameters
$\beta$ and $g=\frac{\omega_{k}}{L}\sqrt{\frac{\hbar}{2m\omega_{0}}}$
are the anharmonic and coupling constants due to radiation pressure,
respectively. The first two terms of the Hamiltonian (\ref{eq:1})
represent the evolution of harmonic oscillators, the third term is
an anharmonic term which is due to the oscillation of the nonlinear
mirror, and the last term corresponds to the coherent interaction
between the quantized cavity mode and the nonlinear mirror. We assume
that the nonlinearity and the coupling constant due to radiation pressure
are considerably weak compared to frequency $\omega_{m}$, so that
$\beta/\omega_{m}$ and $g/\omega_{m}$ are very small compared to
unity. We further assume that the interaction time $t$ is such that
$\beta t$ and $gt$ are less than unity and terms involving cubes
or higher power in $\beta t$ and $gt$ can safely be neglected. In
the present work, we have considered equilibrium cavity length $L\sim0.04$
m, mass of the movable mirror $m\sim10^{-14}$kg, and frequency of
the oscillating mirror $\omega_{0}\sim10^{4}\,Hz$. This choice of
parameters is consistent with the existing literature \cite{key-4-1}. 

The Hamiltonian (\ref{eq:1}) represents a general form of an optomechanical
system which contains  nonlinearity. If we consider the nonlinearity
constant $\beta=0$, we would obtain a special case of the Hamiltonian
(\ref{eq:1}) that will be exactly solvable and would be mathematically
equivalent to an optomechanics-like system in which Bose-Einstein
condensate (BEC) is trapped inside an optical resonator driven by
the quantized light field \cite{BEC_Prl}. As mentioned in the previous
section, in \cite{BEC_Prl}, the authors have studied the role reversal
between matter wave and quantized light, i.e., matter wave behaves
like the quantized light field and vice versa. With the condition
$\beta=0$, the Hamiltonian (\ref{eq:1}) takes the form 

\begin{equation}
\begin{array}{lcl}
H & = & \hbar\omega_{k}k^{\dagger}k+\hbar\omega_{0}a^{\dagger}a-\hbar g(k^{\dagger}ka^{\dagger}+{\rm H.c.}).\end{array}\label{eq:2-1}
\end{equation}
where $\omega_{m}$ reduces to $\omega_{0}$ in Eq. (\ref{eq:1})
as $\beta=0$, and the third term vanishes. The reduced Hamiltonian
(\ref{eq:2-1}) corresponds to a simple form of a Fabry\textendash Perot
cavity with one movable mirror, and it also describes a trapped BEC
inside the cavity resonator, where $\omega_{k}$ represents the frequency
of the mater wave ($k^{\dagger}k$) originated from the center of
mass of the trapped BEC, and $\omega_{0}$ corresponds to the cavity-pump
detuning \cite{BEC_Prl}.\textcolor{red}{{} }In the BEC system, the
role of the movable mirror (in the Fabry-Perot-type sytems) is played
by the ultracold gas, which is the trapped BEC in this particular
case. The analogue of optomechanical coupling is taken as the coupling
between the trapped BEC mode and the cavity field detuning. In what
follows, the analytic and numerical results for the BEC system is
obtained with cavity pump detuning $\omega_{k}=10^{4}\,Hz,$ the renormalized
ground-state center-of-mass frequency of the trapped BEC $\omega_{0}=10^{5}\,Hz$
and coupling constant $g=0.0002\times\omega_{0}\,Hz$. A similar situation
has been investigated experimentally in various works \cite{BEC_SCience,BEC_Nature}.
It may be noted that in Eq. (\ref{eq:2-1}), the coupling can be visualized
to be originated from the radiation pressure from a \emph{massive}
Schrodinger field driving an optical oscillator. This is in contrast
to the usual description that considers that a real radiation pressure
from a massless optical field drives a mechanical oscillator \cite{BEC_Prl}.

\section{the solution\label{sec3:the-solution}}

Our intention is to solve the Hamiltonian (\ref{eq:1}) in the Heisenberg
picture. Generally, it is difficult to obtain closed form analytic
solution in the Heisenberg picture due to the presence of non-commuting
operators. Perhaps, a more accurate numerical solution may be possible
for the present problem. However, in order to obtain a much more physical
insight into the system, we prefer an approximate analytic solution
using a perturbative technique known as Sen-Mandal technique (\cite{AP_BEC-1,bsen1,KT,KT-2,KT-3,Nasir,nasir-sq,nasir-phase}
and references therein). In order to obtain the closed form analytic
expressions for the time evolution of the operators under weak nonlinearity
and low radiation pressure coupling, we start with the following Heisenberg's
equations of motion that are obtained from the Hamiltonian (\ref{eq:1}) 

\begin{equation}
\begin{array}{lcl}
\dot{a} & = & -i\omega_{m}a-2i\beta a^{\dagger}a^{2}+igk^{\dagger}k,\\
\dot{k} & = & -i\omega_{k}k+igk(a^{\dagger}+a),
\end{array}\label{eq:2}
\end{equation}
where the over-dots correspond to the differentiation with respect
to time $t$, and the coupling constant $g$ and nonlinearity constant
$\beta$ are assumed to be real. The coupled nonlinear differential
equations in Eq. (\ref{eq:2}) involving the operators $a$ and $k$
are not exactly solvable. However, perturbative analytic solution
for the time evolution of operators $a$ and $k$ can be obtained
in terms of $t,\,a(0)$, $a^{\dagger}(0)$, $k(0)$, and $k^{\dagger}(0)$
and other parameters of the system by using Sen-Mandal approach, which
has already been successfully used to study various physical systems
that lead to Heisenberg's equations of motion involving the coupled
nonlinear differential equations \cite{AP_BEC-1,bsen1,KT,KT-2,KT-3,Nasir,nasir-sq,nasir-phase}.
We have two special situations, where these differential equations
(\ref{eq:2}) and hence the model Hamiltonian (\ref{eq:1}) offers
an exact analytic solution. The first one is for $g=0$ (i.e., when
there is no coupling), two differential equations (\ref{eq:2}) are
completely decoupled and the system behaves like two independent oscillators.
In this case, for mode $a(t)$, the initial field frequency $\omega_{0}$
gets modified to $\omega_{m}$ due to the presence of a Kerr-type
nonlinearity involving the nonlinear constant $\beta$. The second
one is $\beta=0$ (i.e., when the mirror oscillates linearly), the
modes $a(t)$ and $k(t)$ are the system of coupled harmonic oscillators,
and are useful for the investigation of nonclassical properties of
the radiation fields. The presence of both the nonlinearity constant
$\beta$ and the coupling constant $g$ makes the coupled equations
(\ref{eq:2}) unsolvable in closed analytic forms. In the present
investigation, we shall keep ourselves confined to the solutions up
to the second orders in $\beta t$ and $gt$, which in our opinion
is reasonable enough to deal with the system described above and all
the related physical problems. In what follows, we will also establish
the consistency of our analytic solution using a numerical solution
of the time dependent Schr{\"o}dinger equation for the same system.
Following Sen-Mandal perturbative approach, the assumed approximated
trial solutions for the operators of the oscillating mirror and cavity
modes up to second order in$\beta t$ and $gt$ can be written as

\begin{widetext}

\begin{equation}
\begin{array}{lcl}
a(t) & \simeq & f_{1}a(0)+f_{2}a^{\dagger}(0)a^{2}(0)+f_{3}k^{\dagger}(0)k(0)+f_{4}a^{\dagger}(0)a(0)a^{\dagger}(0)a^{2}(0)+f_{5}k^{\dagger}(0)k(0)a^{\dagger}(0)a(0)+f_{6}k^{\dagger}(0)k(0)a^{2}(0),\\
k(t) & \simeq & h_{1}k(0)+h_{2}k(0)a^{\dagger}(0)+h_{3}k(0)a(0)+h_{4}k(0)a^{\dagger2}(0)a(0)+h_{5}k(0)a^{\dagger}(0)a^{2}(0)+h_{6}k(0)a^{\dagger2}(0)+h_{7}k(0)a^{2}(0)\\
 & + & h_{8}k(0)a^{\dagger}(0)a(0)+h_{9}k(0)a(0)a^{\dagger}(0)+h_{10}k(0)k^{\dagger}(0)k(0).
\end{array}\label{eq:4}
\end{equation}
\end{widetext}The notations $\simeq$ in the above equations and
the later part of the paper is used to indicate that the terms beyond
the second order in $\beta t$ and $gt$ are neglected from the right
hand side of the corresponding equations. In the rest of the paper,
we will simply write $a$ and $k$ instead of $a(0)$ and $k(0)$.
The similar notation will be adopted for the corresponding creation
operators. The assumed solution given in Eq. (\ref{eq:4}) would not
be considered complete unless we evaluate the functional form of the
time dependent parameters $f_{i}$ and $h_{i}$. Therefore, after
differentiating Eq. (\ref{eq:4}) with respect to time $t$ and substituting
that in Eq. (\ref{eq:2}), we obtain a set of coupled first order
differential equations involving $f_{i}$ and $h_{i}$. The time dependent
coefficients $f_{i}$ and $h_{i}$ are actually obtained from the
dynamics involving Eqs. (\ref{eq:2}) and (\ref{eq:4}) under a set
of initial conditions . Here, we consider $f_{1}=h_{1}=1$ and $f_{i}=h_{i}=0$
(for $i=2,3,4,\cdots$) as the initial conditions, i.e., at $t=0$.
Under these initial conditions the corresponding solutions for the
time dependent coefficient $f_{i}$s are obtained as 

\begin{equation}
\begin{array}{lcl}
f_{1} & = & e^{-i\omega_{m}t},\\
f_{2} & = & -2i\beta tf_{1},\\
f_{3} & = & \frac{g}{\omega_{m}}\left[1-f_{1}\right],\\
f_{4} & = & -i\beta tf_{2},\\
f_{5} & = & \frac{4g\beta}{\omega_{m}^{2}}f_{1}\left[-e^{i\omega_{m}t}+i\omega_{m}t+1\right],\\
f_{6} & = & \frac{2g\beta}{\omega_{m}^{2}}f_{1}\left[e^{-i\omega_{m}t}+i\omega_{m}t-1\right].
\end{array}\label{eq:5}
\end{equation}

On the other hand, to obtain the solution for the quantized cavity
mode $k(t)$, we need to find out the solutions for the time dependent
coefficients $h_{i}$s. Under the initial boundary condition mentioned
previously, these are obtained as

\begin{equation}
\begin{array}{lcl}
h_{1} & = & e^{-i\omega_{k}t},\\
h_{2} & = & h_{1}f_{3}/f_{1},\\
h_{3} & = & h_{1}f_{3},\\
h_{4} & = & h_{1}f_{6}/f_{1}^{2},\\
h_{5} & = & h_{1}f_{5}/2,\\
h_{6} & = & h_{1}f_{3}^{2}/(2f_{1}^{2}),\\
h_{7} & = & h_{1}f_{3}^{2}/2,\\
h_{8} & = & \frac{g}{2\beta}h_{1}f_{6}/f_{1},\\
h_{9} & = & -\frac{g}{4\beta}h_{1}h_{5}/f_{1},\\
h_{10} & = & f_{8}-f_{9}.
\end{array}\label{eq:6}
\end{equation}

The coefficients $f_{1}$ and $h_{1}$ in Eqs. (\ref{eq:5}) and (\ref{eq:6})
are due to the free evolution terms which correspond to the harmonic
oscillators, and the rest of the coefficients are due to the nonlinearity
and/or coupling constants. The solutions of the coefficients $f_{i}$s
and $h_{i}$s complete the operator solutions of $a(t)$ and $k(t)$.
The operators $a(t)$ and $k(t)$ (i.e., corresponding to the movable
mirror mode and quantized cavity mode) are the bosonic operators,
and hence they must obey the bosonic commutation relations. It may
be noted that the obtained solutions are verified to obey the equal
time commutation relations (ETCR), i.e., $\left[a(t),a^{\dagger}(t)\right]=\left[k(t),k^{\dagger}(t)\right]=1$,
while all other possible commutations vanish as required. From the
above equations, it can be seen that the operators $a(t)$ and $k(t)$
commute with each other. As a consequence of this commutation relation,
in the rest of the paper, we can compute various moments of these
operators without being worried about the intermodal commutation relation
and the corresponding operator ordering. In what follows, we use the
closed form analytic expressions of $a(t)$ and $k(t)$ obtained here
to investigate the temporal evolution of lower and higher-order entanglement
and different nonclassical properties of the system, using various
moment based criteria. To begin with, we look at the possibility of
observing squeezed state in the next section. 

\section{squeezing\label{sec4:squeezing}}

In this section, we investigate the possibilities of observing squeezing,
higher-order squeezing and intermodal squeezing of the quadratures.
In general, the quadrature operators of the various field modes are
defined as 

\begin{equation}
\begin{array}{lcl}
X_{j} & = & \frac{1}{2}[j(t)+j^{\dagger}(t)],\\
Y_{j} & = & -\frac{i}{2}[j(t)-j^{\dagger}(t)],
\end{array}\label{eq:-1}
\end{equation}
where $X_{j}$ $(Y_{j})$ is the quadrature operators of the corresponding
modes with $j\in\left\{ a,\,k\right\} $. The fluctuation of these
quadratures obey the famous uncertainty relation, and its minimum
value corresponds to the minimum uncertainty state (MUS). The fluctuation
of the minimum uncertainty state gives rise to the standard quantum
limit (SQL), and hence the zero point fluctuation (ZPF). The quasi-classical
states, i.e., coherent states are the examples of the MUS in which
uncertainties in both of the quadratures are $\frac{1}{2}$ (in dimensionless
unit). If one of the quadrature fluctuation goes below the SQL for
a quantum state, then the corresponding state is called squeezed state.
In such a situation, in order to respect the Heisenberg uncertainty
relation the other quadrature fluctuation must be greater than the
SQL and hence, the simultaneous squeezing of both of the quadratures
is not allowed, i.e., squeezing in one of the quadrature components
automatically prohibits the same in the conjugate quadrature. The
squeezing of the quadrature $X_{j}$ $(Y_{j})$ is obtained if the
second order variance $(\Delta X_{j})^{2}<\frac{1}{4}$ $\left((\Delta Y_{j})^{2}<\frac{1}{4}\right)$.
In order to calculate $(\Delta X_{j})^{2}$ and $(\Delta Y_{j})^{2}$,
we assume that the cavity and oscillating mirror modes are initially
in the coherent state. Therefore, the composite initial state is the
product of the states $|\alpha_{1}\rangle$ and $|\alpha_{2}\rangle$,
which are the eigenket of the operators $a$ and $k$. Thus, the initial
composite state is

\begin{equation}
|\psi(0)\rangle=|\alpha_{1}\rangle\otimes|\alpha_{2}\rangle.\label{eq:initial state}
\end{equation}
Explicitly, the initial state is separable, and the operators $a$
and $k$ obey the eigenvalue equations

\begin{equation}
\begin{array}{lcl}
a(0)|\psi(0)\rangle & = & \alpha_{1}|\alpha_{1}\rangle\otimes|\alpha_{2}\rangle,\\
k(0)|\psi(0)\rangle & = & \alpha_{2}|\alpha_{1}\rangle\otimes|\alpha_{2}\rangle,
\end{array}\label{eq:10-1}
\end{equation}
where $\alpha_{1}=|\alpha_{1}|e^{-i\theta}$ and $\alpha_{2}=|\alpha_{2}|e^{-i\phi}$
are the complex quantities. The parameters $|\alpha_{1}|^{2}$ and
$|\alpha_{2}|^{2}$ represent the number of excitation present in
the modes $a$ and $k$, respectively. Also, the quantities $\theta$
and $\phi$ are the phase angles of the vibrating mirror and quantized
cavity modes, respectively. In what follows, we will show that varying
phase angle $\theta$ we can control the depth (values) of nonclassicality
parameters. Similarly, it is expected that by controlling $\phi$,
we would be able to control the values of the nonclassicality parameters.
However, in all the plots reported here, id assumed that $\phi=0.$

\begin{widetext}

\begin{figure}[h]
\begin{centering}
\begin{tabular}{ccc}
\includegraphics[scale=0.4]{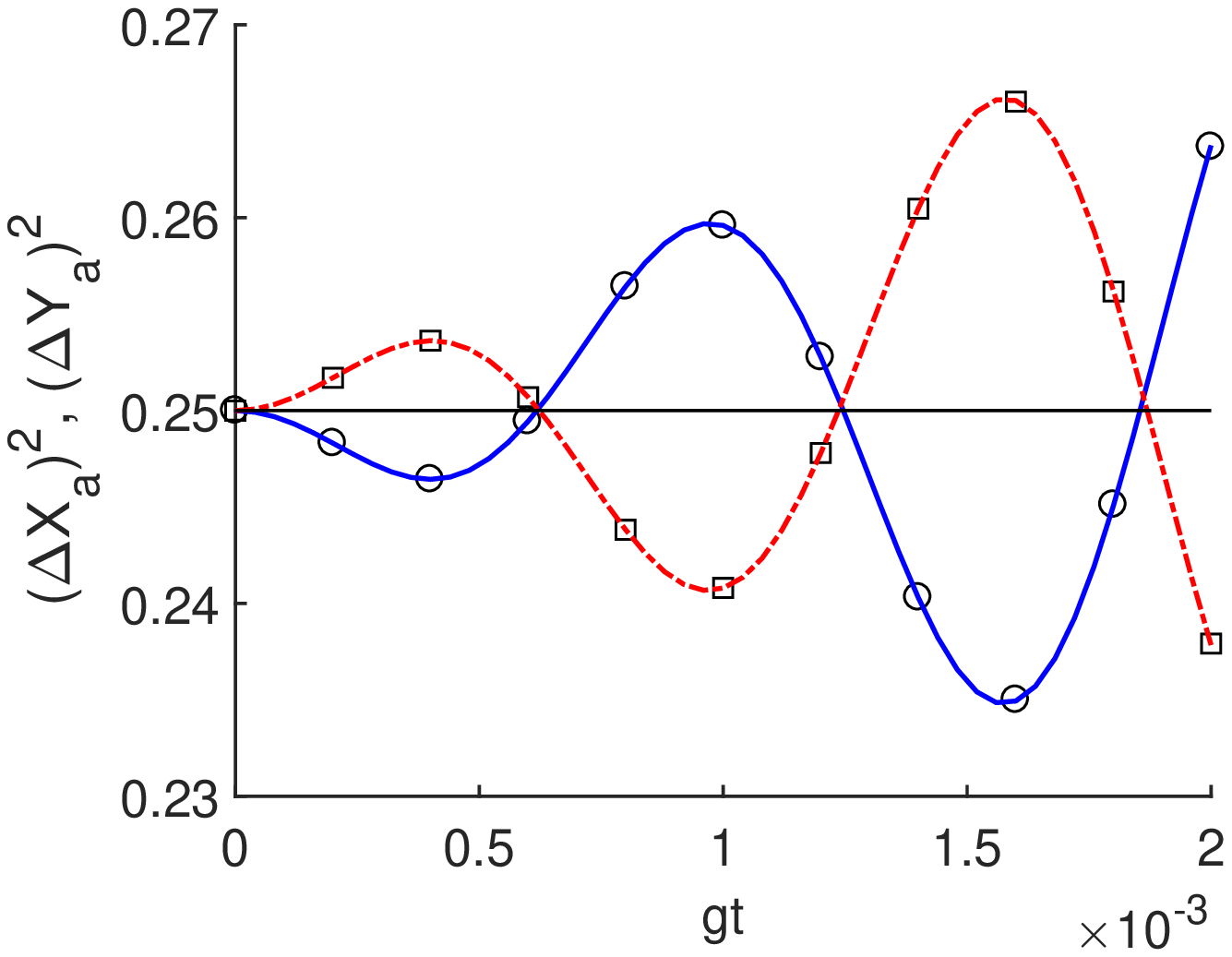} & \includegraphics[scale=0.4]{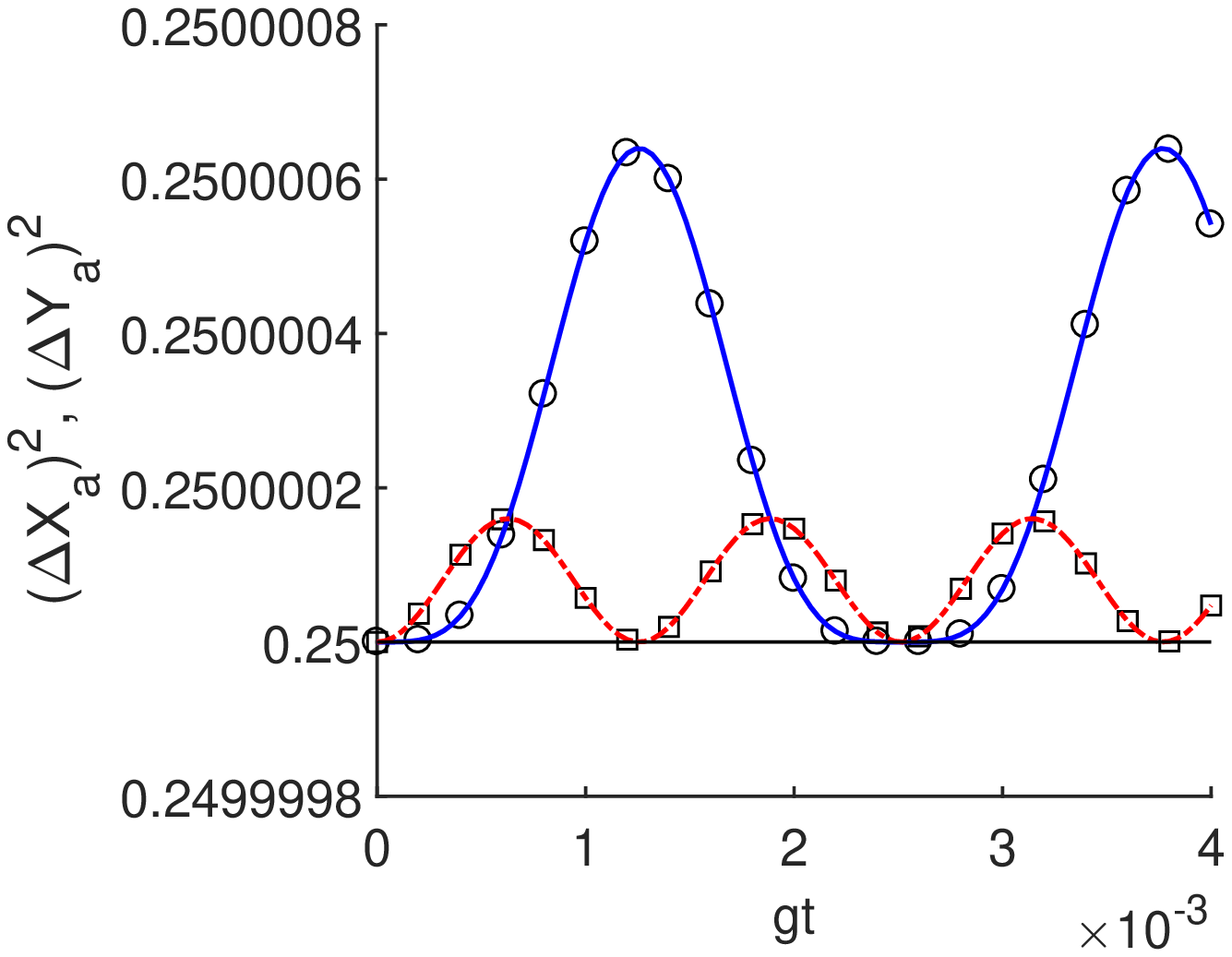} & \includegraphics[scale=0.4]{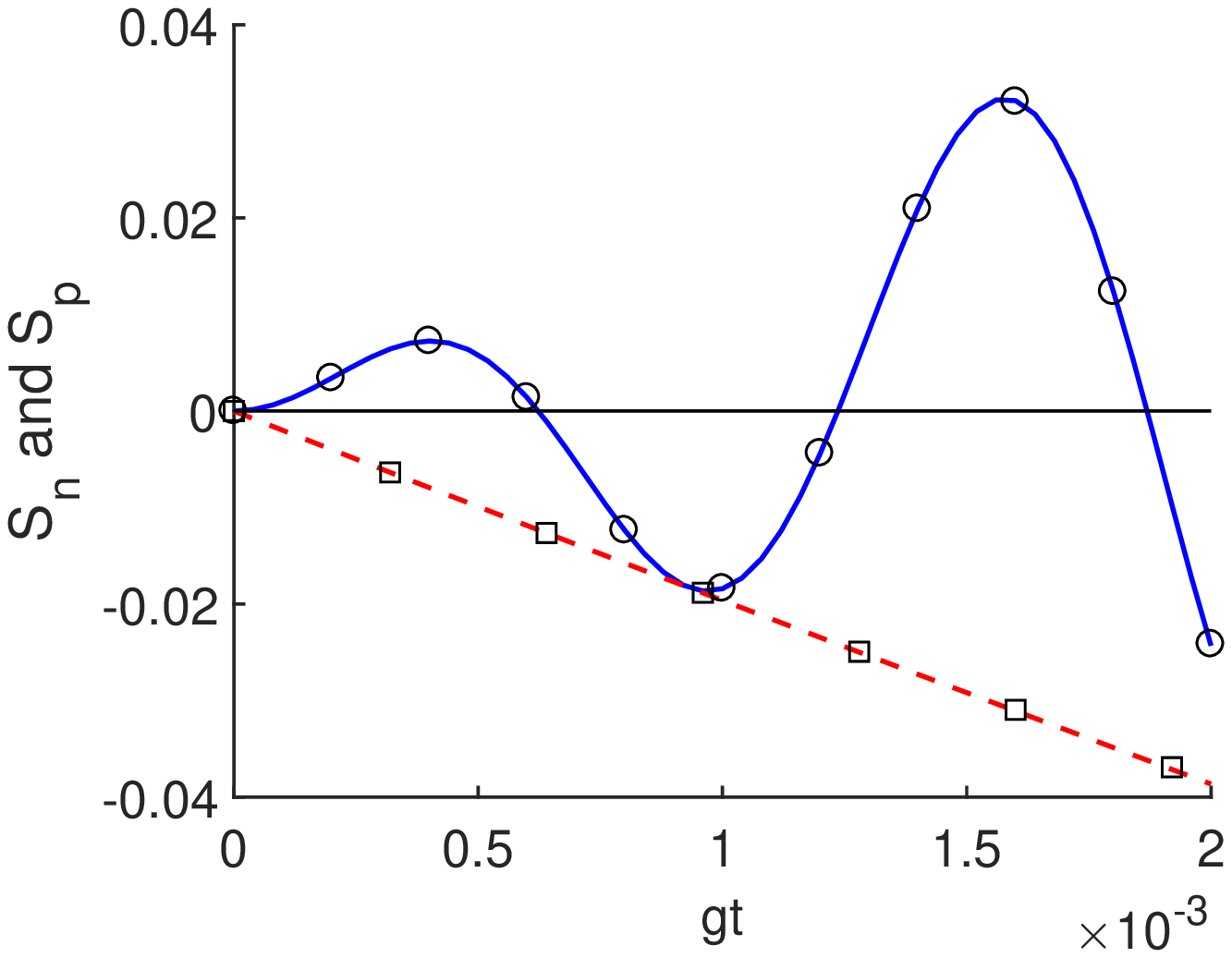}\tabularnewline
(a) & (b) & (c)\tabularnewline
\includegraphics[scale=0.4]{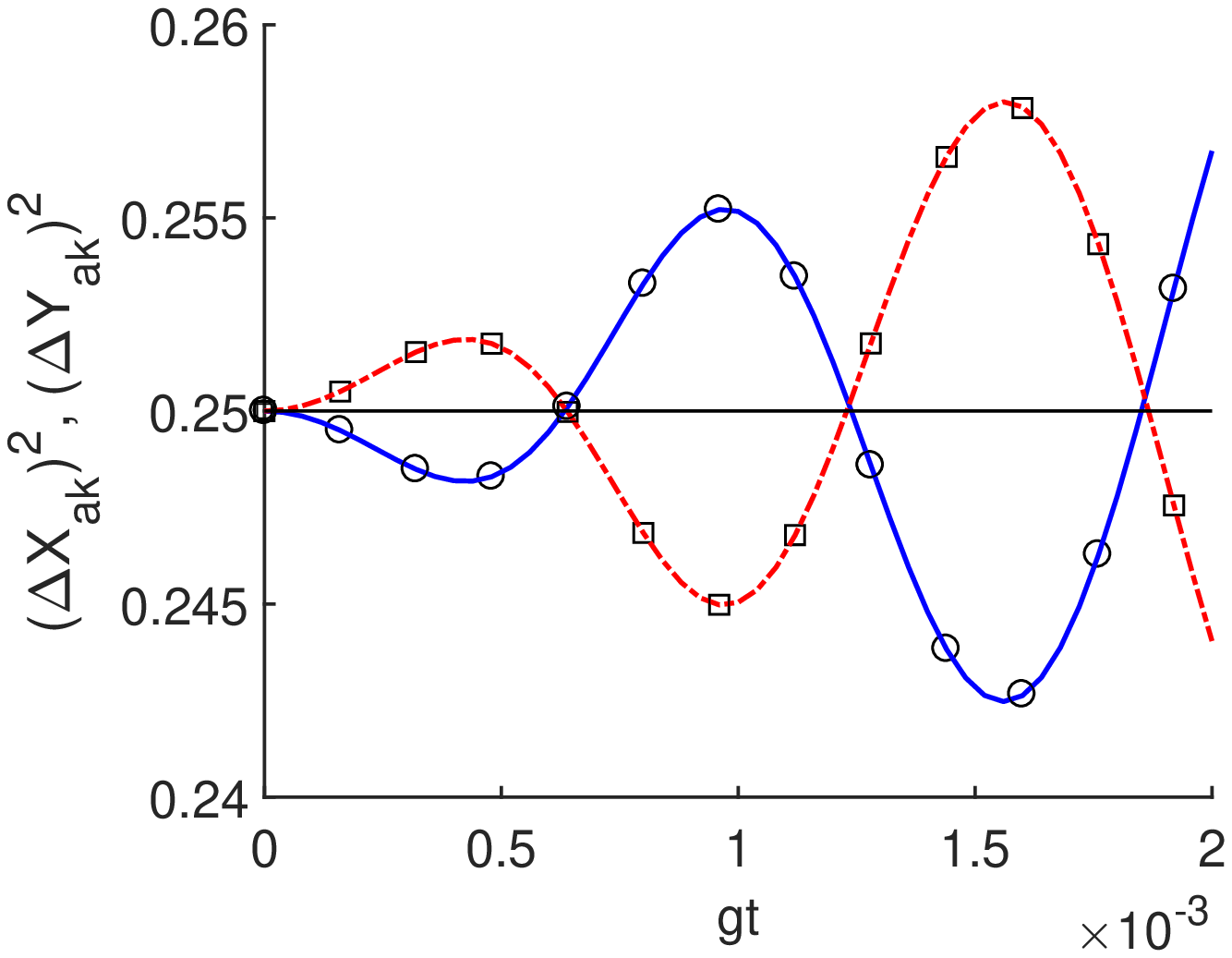} & \includegraphics[scale=0.4]{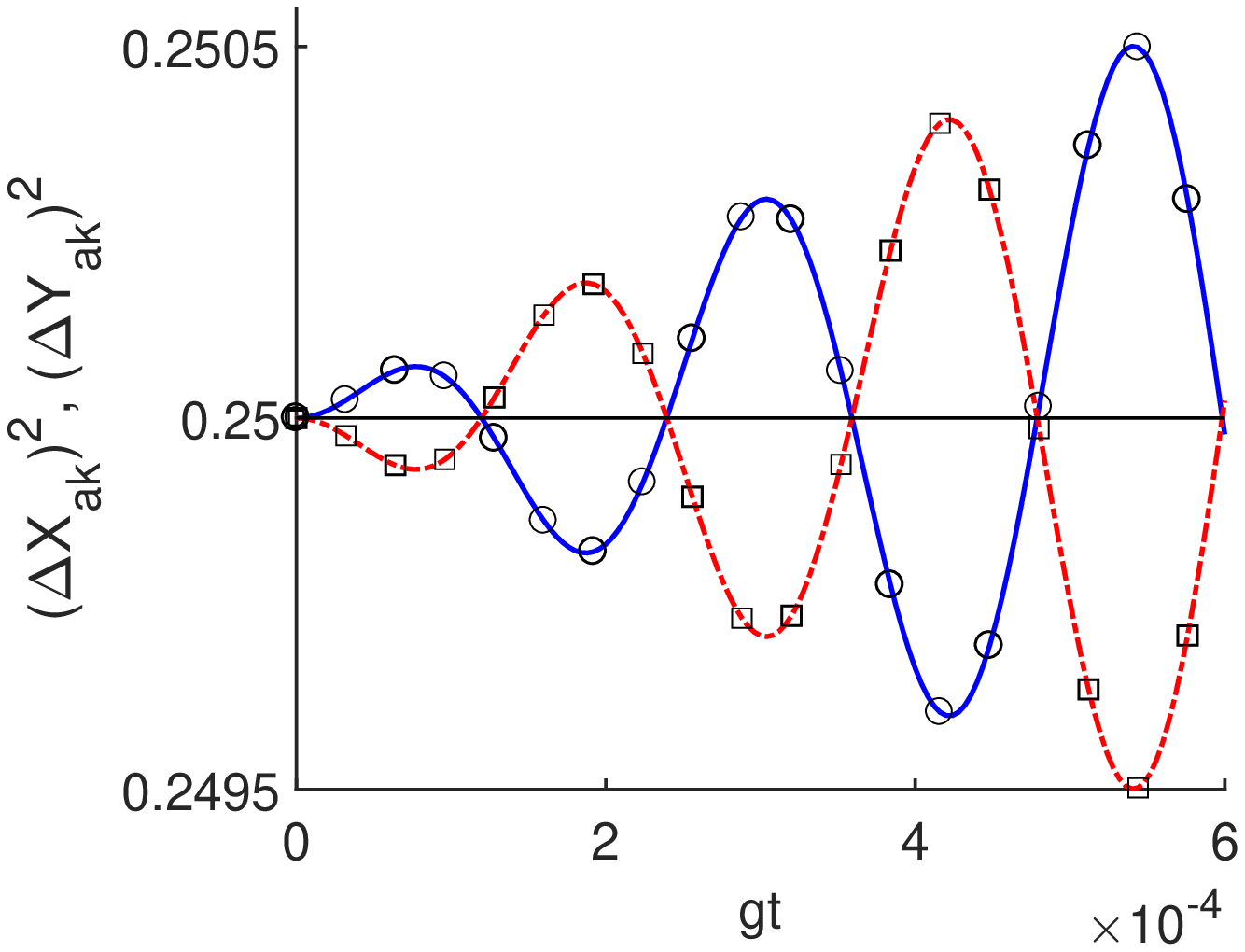} & \includegraphics[scale=0.4]{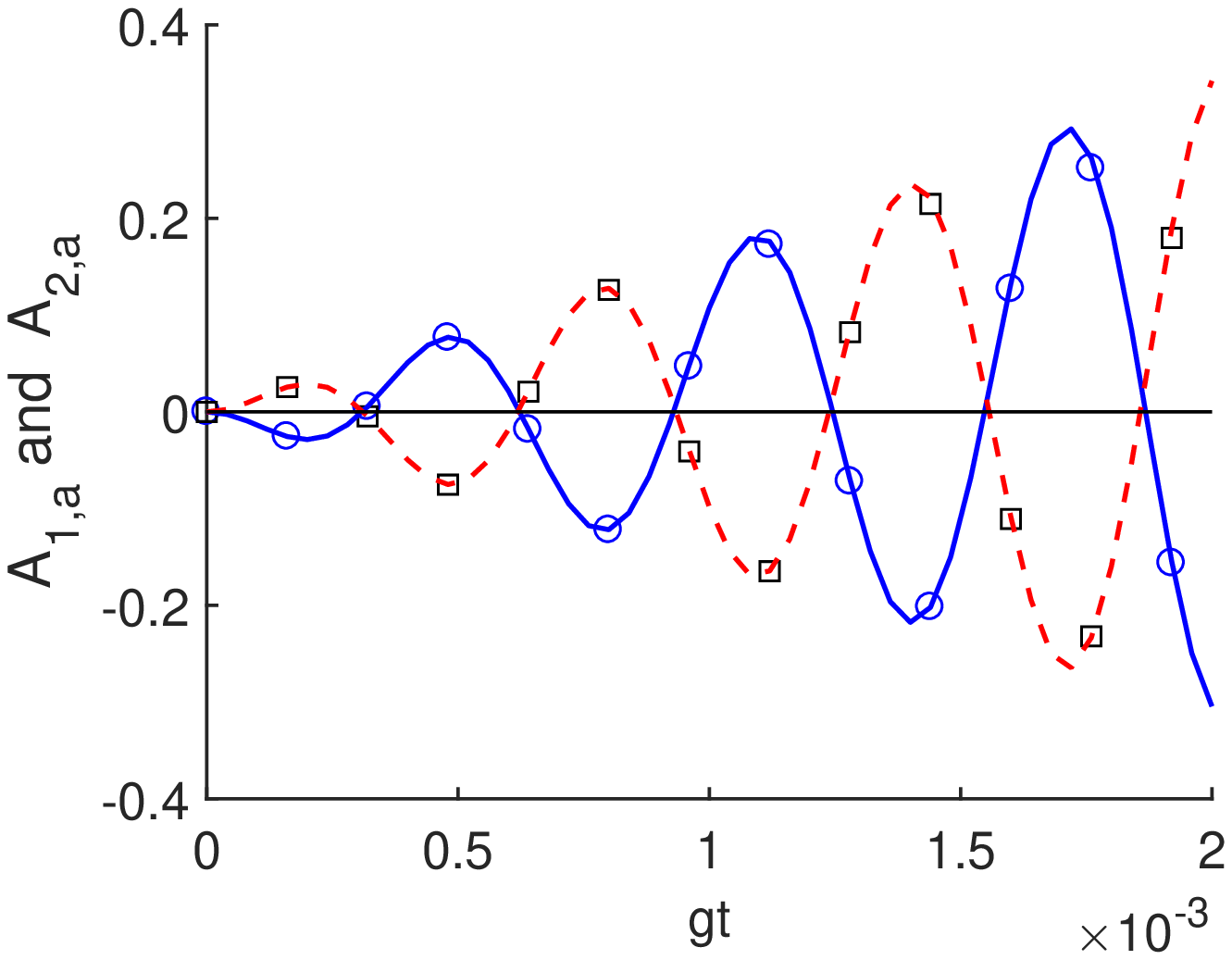}\tabularnewline
(d) & (e) & (f)\tabularnewline
\end{tabular}
\par\end{centering}
\caption{\label{fig:squeezing}(Color online) Variation of single mode lower
and higher-order quadrature squeezing, intermodal, principal and normal
squeezing for optomechanical system (in (a), (c), (d), and (f)), and
optomechanics-like system (in (b), (e)). (a) Illustration of quadrature
squeezing with rescaled time in mode $a$. The analytic results for
$(\varDelta X_{a})^{2}$ and $(\varDelta Y_{a})^{2}$ correspond to
solid (blue) and dashed-dotted (red) curves, respectively. (b) The
presence of quadrature squeezing for mode $a$ in optomechanics-like
system, where solid (blue) and dashed (red) lines correspond to $(\Delta X_{a})^{2}$
and $(\Delta Y_{a})^{2}$, respectively. (c) Illustration of principal
and normal squeezing through dotted (red) and smooth (blue) lines,
respectively. (d) Variation of intermodal squeezing with rescaled
time via solid (blue) and dashed-dotted (red) curves corresponding
to $(\varDelta X_{ak})^{2}$ and $(\varDelta Y_{ak})^{2}$. Also,
(e) intermodal squeezing between modes $a$ and $k$ in trapped-BEC
system is shown using solid (blue) and dashed (red) lines corresponding
to $(\Delta X_{ak})^{2}$ and $(\Delta Y_{ak})^{2}$, respectively.
(f) Illustration of $l$th powered quadrature squeezing in mode $a$
for $l=2$, where the solid (blue) and dashed (red) curves represent
the analytic results of $A_{1,a}$ and $A_{2,a}$, respectively. For
all the results of optomechanical system (i.e., (a), (c), (d) and
(f)), we have considered $|\alpha_{1}|^{2}=4$, $|\alpha_{2}|^{2}=1$,
while $|\alpha_{1}|^{2}=2,$ $|\alpha_{2}|^{2}=4$ for optomechanics-like
BEC system (i.e., (b), (e)), and $\theta=\phi=0$ for all the figures,
unless stated otherwise. In this figure, and all the following figures,
the plotmarkers (i.e., circles, squares, diamonds and stars) are used
to represent the values obtained from the numerical solution for the
corresponding quantity. Also, all the quantities that are shown in
all the figures are dimensionless. }
\end{figure}

\end{widetext}

Now, in terms of the initial composite coherent state (\ref{eq:initial state}),
we calculate the second order variance of quadratures $X_{j}$ and
$Y_{j}$. Therefore, using Eqs. (\ref{eq:4})-(\ref{eq:10-1}), we
have

\begin{equation}
\begin{array}{lcl}
\left[\begin{array}{c}
(\varDelta X_{a})^{2}\\
(\varDelta Y_{a})^{2}
\end{array}\right] & = & \frac{1}{4}\left[1+2(|f_{2}|^{2}|\alpha_{1}|^{4}+|f_{3}|^{2}|\alpha_{2}|^{2})\right.\\
 & \pm & \left\{ f_{1}f_{2}\alpha_{1}^{2}+f_{1}f_{4}(2|\alpha_{1}|^{2}\alpha_{1}^{2}+\alpha_{1}^{2})\right.\\
 & + & f_{1}f_{5}|\alpha_{1}|^{2}\alpha_{1}+f_{2}^{2}2|\alpha_{1}|^{2}\alpha_{1}^{2}\\
 & + & \left.\left.f_{3}^{2}|\alpha_{2}|^{2}+{\rm c.c.}\right\} \right],
\end{array}\label{eq:sqz}
\end{equation}
where ${\rm c.c.}$ stands for the complex conjugate. The upper and
lower signs ($+$ and $-$ sign) in the right hand side of Eq. (\ref{eq:sqz})
correspond to $(\varDelta X_{a})^{2}$ and $(\varDelta Y_{a})^{2}$,
respectively. It is clear from the above expression that for $t=0$,
the variances reduce to their minimum values $(\varDelta X_{a})^{2}=$
$(\varDelta Y_{a})^{2}=\frac{1}{4}$, hence belong to the coherent
state as expected, which is clearly depicted in Fig. \ref{fig:squeezing}
a. In which, one can observe that at $t=0$ curves start from $0.25$
and with increase of rescaled interaction time, the values of $(\varDelta X_{a})^{2}$
or $(\varDelta Y_{a})^{2}$ show an oscillatory behavior. From Fig.
\ref{fig:squeezing} a, it's clear that at certain times $(\varDelta X_{a})^{2}$$\left((\varDelta Y_{a})^{2}\right)$
goes below the SQL and thus indicate the exisitence of squeezing in
$X_{a}\,(Y_{a})$ quadrature. Additionally, with $\beta=0$, the variances
will correspond to the BEC trapped optomechanics-like system described
by Eq. (\ref{eq:2-1}), and the corresponding result is shown in Fig.
\ref{fig:squeezing} b, where we failed to obtain squeezing in any
quadrature corresponding to mode $a$. For obtaining the figures for
the trapped BEC system, we have chosen the cavity pump detuning $\omega_{k}=10^{4}\,Hz,$
the renormalized ground-state center-of-mass frequency of the trapped
BEC $\omega_{0}=10^{5}\,Hz$, and coupling constant $g=0.0002\times\omega_{0}\,Hz$.
Further, as mentioned previously, we obtained the quadrature squeezing
in mode $a$ solving the system Hamiltonian (\ref{eq:1}) in Schr{\"o}dinger
picture using matrix forms of various operators with the same initial
state. Interestingly, the quadrature squeezing obtained using perturbative
solution is observed to match exactly with that obtained from exact
numerical solution (which is shown as circles and squares in Fig.
\ref{fig:For-the-BEC} a). Similarly, we have verified all the results
reported in the present work using numerical solution, which are shown
using various plotmarkers.

An alternate definition of quadrature squeezing was given by Luks
et al., \cite{luks1 et al} by considering the geometrical (elliptical)
representation of variances. Further advancement of the geometrical
representation of squeezed state happened in the works of Loudon \cite{Loudon1},
who represented the quadrature variance in terms of the Booth's elliptical
lemniscate. According to Luks et al., \cite{luks1 et al}, the standard
definition of squeezing (\textit{normal squeezing}) is 

\begin{equation}
S_{n}=\langle\varDelta a^{\dagger}\varDelta a\rangle-Re\langle(\varDelta a)^{2}\rangle<0,\label{eq:1-1}
\end{equation}
and the definition of\textit{ principal squeezing} is given by 

\begin{equation}
S_{p}=\langle\varDelta a^{\dagger}\varDelta a\rangle-|\langle(\varDelta a)^{2}\rangle|<0.\label{eq:2-2}
\end{equation}

The expectation values are calculated in terms of the initial composite
coherent state. The analytic expressions of the normal and principal
squeezing are obtained using Eqs. (\ref{eq:4}), (\ref{eq:1-1}) and
(\ref{eq:2-2}) as follows

\begin{widetext}

\begin{equation}
\begin{array}{lcl}
S_{n} & = & |f_{2}|^{2}|\alpha_{1}|^{4}+|f_{3}|^{2}|\alpha_{2}|^{2}-Re\left[f_{1}f_{2}\alpha_{1}^{2}+f_{1}f_{4}(2|\alpha_{1}|^{2}\alpha_{1}^{2}+\alpha_{1}^{2})+f_{1}f_{5}|\alpha_{1}|^{2}\alpha_{1}+2f_{2}^{2}|\alpha_{1}|^{2}\alpha_{1}^{2}+f_{3}^{2}|\alpha_{2}|^{2}\right]\end{array}\label{eq:13}
\end{equation}
and

\begin{equation}
\begin{array}{lcl}
S_{p} & = & |f_{2}|^{2}|\alpha_{1}|^{4}+|f_{3}|^{2}|\alpha_{2}|^{2}-|\left(f_{1}f_{2}\alpha_{1}^{2}+f_{1}f_{4}(2|\alpha_{1}|^{2}\alpha_{1}^{2}+\alpha_{1}^{2})+f_{1}f_{5}|\alpha_{1}|^{2}\alpha_{1}+2f_{2}^{2}|\alpha_{1}|^{2}\alpha_{1}^{2}+f_{3}^{2}|\alpha_{2}|^{2}\right)|,\end{array}
\end{equation}

\end{widetext}respectively. The variations of $S_{n}$ and $S_{p}$
are depicted in Fig. \ref{fig:squeezing} c, where $S_{p}$ is always
negative, but $S_{n}$ is positive in some regions. The negative values
of $S_{n}$ and $S_{p}$ is the clear witness of the presence of quadrature
squeezing. 

\subsection{Intermodal squeezing}

In this subsection, we investigate the intermodal squeezing between
the quantized cavity mode and the phonon mode that correspond to the
vibrating mirror. The two-mode quadrature operators involving the
modes $a$ and $k$ are defined as follows \cite{IntSq} 
\begin{equation}
\begin{array}{lcl}
X_{ak} & = & \frac{1}{2\sqrt{2}}\{(a+a^{\dagger})+(k+k^{\dagger})\},\\
Y_{ak} & = & -\frac{i}{2\sqrt{2}}\{(a-a^{\dagger})+(k-k^{\dagger})\},
\end{array}\label{eq:msqz}
\end{equation}
where the variances of quadratures $X_{ak}$ and $Y_{ak}$ obey the
Heisenberg's uncertainty relation. Consequently, whenever the variance
in one of the quadratures goes below $\frac{1}{4}$, it demonstrates
squeezing in corresponding two-mode quadrature. Using Eqs. (\ref{eq:4})-(\ref{eq:6}),
(\ref{eq:initial state}), (\ref{eq:10-1}) and (\ref{eq:msqz}) we
obtain the expressions for the variances of the intermodal quadratures
as

\begin{widetext}

\begin{equation}
\begin{array}{lcl}
\left[\begin{array}{c}
(\varDelta X_{ak})^{2}\\
(\varDelta Y_{ak})^{2}
\end{array}\right] & = & \frac{1}{8}[2+2(|f_{2}|^{2}|\alpha_{1}|^{4}+|f_{3}|^{2}|\alpha_{2}|^{2})+2|h_{2}|^{2}|\alpha_{2}|^{2}+2f_{2}h_{2}^{*}\alpha_{2}^{*}\alpha_{1}^{2}\\
 & \pm & \left\{ f_{1}f_{2}\alpha_{1}^{2}+f_{1}f_{4}(2|\alpha_{1}|^{2}\alpha_{1}^{2}+\alpha_{1}^{2})+f_{1}f_{5}|\alpha_{1}|^{2}\alpha_{1}+2f_{2}^{2}|\alpha_{1}|^{2}\alpha_{1}^{2}+f_{3}^{2}|\alpha_{2}|^{2}\right.\\
 & + & \left.h_{1}h_{10}\alpha_{2}^{2}+h_{2}h_{3}\alpha_{2}^{2}\right\} \pm2\left\{ f_{1}h_{2}\alpha_{2}+2(f_{1}h_{4}+f_{2}h_{2})\alpha_{2}|\alpha_{1}|^{2}+f_{1}h_{5}\alpha_{2}\alpha_{1}^{2}\right.\\
 & + & \left.2f_{1}h_{6}\alpha_{2}\alpha_{1}^{*}+(f_{1}h_{8}+f_{1}h_{9})\alpha_{2}\alpha_{1}+{\rm c.c.}\right\} ],
\end{array}\label{eq:18-1}
\end{equation}
\end{widetext}where upper and lower signs in the left hand side correspond
to the variances of the compound mode quadratures $(\varDelta X_{ak})^{2}$
and $(\varDelta Y_{ak})^{2}$, respectively. The variations of the
intermodal quadrature squeezing is shown in Fig. \ref{fig:squeezing}
d, where one can clearly see an oscillating behavior. It is also clear
that the simultaneous squeezing is not feasible for the intermodal
squeezing as well, i.e., as restricted by the Heisenberg's uncertainty
principle. In order to study the possibility of observing intermodal
squeezing, in the trapped BEC system, we can use $\beta=0$ in Eq.
(\ref{eq:18-1}), where we observed the presence of intermodal squeezing
between modes $a$ and $k$. The variation of intermodal squeezing
in the BEC system with rescaled time is illustrated in Fig. \ref{fig:squeezing}
e. Here, it's important to note that the quadrature squeezing in the
cavity mode is absent in the optomechanics-like system, but intermodal
squeezing can be observed in this system (cf. Fig. \ref{fig:squeezing}
d and e).

\subsection{Higher -order squeezing}

The idea of the higher-order squeezing originated from the work of
Hong and Mandel \cite{HOngMandal} who generalized the concept of
the normal order squeezing. They introduced higher-order squeezing
involving higher-order moments of quadrature variables. On the other
hand, Hillery proposed a notion of higher-order squeezing by computing
the variance of field quadratures which were higher-order in amplitude
\cite{HilleryHOS}, i.e., amplitude powered squeezing. In order to
investigate the higher-order squeezing, here we follow the idea of
Hillery \cite{HilleryHOS}. In which, $l$th powered quadrature variables
are defined as 

\begin{equation}
\begin{array}{lcl}
Y_{1,a} & = & \frac{a^{l}+a^{\dagger l}}{2},\\
Y_{2,a} & = & -i\frac{a^{l}-a^{\dagger l}}{2}.
\end{array}\label{eq:19}
\end{equation}
Here, we can see that the quadrature variables $Y_{1,a}$ and $Y_{2,a}$
do not commute which gives us an uncertainty relation, and consequently a criterion for higher-order squeezing as 
\begin{equation}
A_{i,a}=\left(\Delta Y_{i,a}\right)^{2}-\frac{1}{2}\left|\left\langle \left[Y_{1,a},Y_{2,a}\right]\right\rangle \right|<0,\label{eq:n-th-squeezing}
\end{equation}
where $i\in\{1,2\}.$ In what follows, we have used the Hillery's
criterion for amplitude square squeezing, i.e., in Eq. (\ref{eq:19}),
we choose $l=2$ and reduced the criterion of higher-order squeezing
given in Eq. (\ref{eq:20}) as

\begin{equation}
\begin{array}{lcl}
A_{i,a} & = & \langle(\Delta Y_{i,a})^{2}\rangle-\langle N_{a}+\frac{1}{2}\rangle<0,\end{array}\label{eq:20}
\end{equation}
where $N_{a}$ is the number operator for the mode $a$. We have calculated
the second order variances of the quadrature for $l=2$, involving
the field operator of mode $a$. After a few algebraic steps and using
Eqs. (\ref{eq:4})-(\ref{eq:6}), (\ref{eq:initial state}), (\ref{eq:10-1})
and (\ref{eq:19}), we obtain the following compact results. 

\begin{widetext}

\begin{equation}
\begin{array}{lcl}
\left[\begin{array}{c}
\langle(\Delta Y_{1,a})^{2}\rangle\\
\langle(\Delta Y_{2,a})^{2}\rangle
\end{array}\right] & = & |f_{1}|^{4}(|\alpha_{1}|^{2}+\frac{1}{2})+|f_{3}|^{2}|\alpha_{2}|^{2}\left(2|\alpha_{1}|^{2}+|\alpha_{2}|^{2}+1\right)+2|f_{2}|^{2}|\alpha_{1}|^{6}\\
 & + & \left[|\alpha_{2}|^{2}\alpha_{1}\left\{ f_{1}f_{3}^{*}+\frac{1}{2}f_{1}f_{5}^{*}|\alpha_{1}|^{2}+f_{1}^{2}f_{3}^{*}f_{2}^{*}(2|\alpha_{1}|^{2}+1)+f_{3}^{*}f_{2}\left(3|\alpha_{1}|^{2}+1\right)\right\} \right.\\
 & \pm & \left[f_{1}^{2}\alpha_{1}^{4}\left\{ f_{1}f_{2}+f_{1}f_{4}\left(2|\alpha_{1}|^{2}+3\right)+f_{2}^{2}\left(4|\alpha_{1}|^{2}+\frac{5}{2}\right)\right\} +f_{1}^{2}\left(f_{1}f_{5}+2f_{3}f_{2}\right)|\alpha_{2}|^{2}\alpha_{1}^{3}\right.\\
 & + & \left.\left.f_{1}^{2}f_{3}^{2}|\alpha_{2}|^{2}\alpha_{1}^{2}\right]+{\rm c.c.}\right].
\end{array}\label{eq:21}
\end{equation}
The average value of the excitation number of the field mode $a$
is given by 

\begin{equation}
\begin{array}{lcl}
\langle N_{a}\rangle=\langle a^{\dagger}(t)a(t)\rangle & = & |f_{1}|^{2}|\alpha_{1}|^{2}+f_{1}^{*}f_{3}\alpha_{1}^{*}|\alpha_{2}|^{2}+f_{1}^{*}f_{5}|\alpha_{2}|^{2}|\alpha_{1}|^{2}\alpha_{1}^{*}+f_{1}^{*}f_{6}|\alpha_{2}|^{2}|\alpha_{1}|^{2}\alpha_{1}\\
 & + & f_{2}^{*}f_{3}|\alpha_{2}|^{2}|\alpha_{1}|^{2}\alpha_{1}^{*}+f_{3}^{*}f_{1}|\alpha_{2}|^{2}\alpha_{1}+f_{3}^{*}f_{2}|\alpha_{2}|^{2}|\alpha_{1}|^{2}\alpha_{1}\\
 & + & |f_{3}|^{2}(|\alpha_{2}|^{4}+|\alpha_{2}|^{2})+f_{5}^{*}f_{1}|\alpha_{2}|^{2}|\alpha_{1}|^{2}\alpha_{1}+f_{6}^{*}f_{1}|\alpha_{2}|^{2}|\alpha_{1}|^{2}\alpha_{1}^{*}.
\end{array}\label{eq:22}
\end{equation}

\end{widetext}Substituting the right hand sides of (\ref{eq:22})
and (\ref{eq:21}) in (\ref{eq:20}), and using the functional forms
of $f_{i}$s given in Eq. (\ref{eq:5}), we analyzed the results for
$A_{1,a}$ and $A_{2,a}$, which are not explicitly written here.
Variation of the amplitude powered quadratures with rescaled time
is illustrated in Fig. \ref{fig:squeezing} f, where we can see that
either $A_{1,a}$ or $A_{2,a}$ can be negative at a particular instant
of time, and thus, one of the quadratures of the optomechanical system
studied here always shows higher-order squeezing in $a$ mode.\textcolor{red}{{} }

In brief, in the present section, we have established that the present
systems can be easily employed to generate squeezed states. Further,
the presence of higher-order squeezing in optomechanical system makes
it feasible to detect the squeezing, in case corresponding lower-order
criterion failed to do so. Additionally, compound modes in both the
systems of interest are shown to exhibit intermodal squeezing.

\section{Antibunching\label{sec5:Antibunching}}

The phenomenon in which the probability of getting two or more photons
simultaneously is less than the probability of getting single photons
is called the photon antibunching. Experimentally, correlation of
intensity was first time observed by Hanbury-Brown and Twiss \cite{BrownTwiss}
in their remarkable experiment using an incandescent light source,
which concluded that the photons come together. This phenomenon in
which the photons come together is known as the bunching of photon.
The natural tendency of light source is to emit a cluster of photons
rather than the single one, but there are light sources in which photons
do not come in cluster, and hence the antibunching of photon is observed
\cite{WallsAntibunch}. In order to investigate the photon bunching
and antibunching, we generally use the quantum statistical properties
of the radiation field. For this, we calculate the second order correlation
function for zero time delay which is defined as 

\begin{equation}
\begin{array}{lcl}
g^{2}(0) & = & \frac{\langle a^{\dagger}(t)a^{\dagger}(t)a(t)a(t)\rangle}{\langle a^{\dagger}(t)a(t)\rangle\langle a^{\dagger}(t)a(t)\rangle}\end{array}.\label{eq:23}
\end{equation}

The second order correlation function is a useful mathematical tool
to deal with the quantum statistical properties of the radiation field.
Our interest is to investigate the phonon statistics of the vibrating
mirror mode. Interestingly, phonon also obey the second order correlation
function \cite{PhononStatstic}. Also, from Eq. (\ref{eq:23}), we
can write 

\begin{equation}
\begin{array}{lcccl}
g^{2}(0)-1 & = & \frac{(\Delta N(t))^{2}-\langle N(t)\rangle}{\langle N(t)\rangle^{2}} & = & \frac{D_{a}(1)}{\langle N(t)\rangle^{2}},\end{array}\label{2nd order correlation}
\end{equation}
where 

\begin{equation}
\begin{array}{lcl}
D_{a}(1) & = & (\Delta N(t))^{2}-\langle N(t)\rangle.\end{array}\label{Da}
\end{equation}
The denominator of Eq. (\ref{2nd order correlation}) is always positive.
Therefore, quantum behavior can be solely determined by the parameter
$D_{a}(1)$. Specifically, for $D_{a}(1)=0$ (i.e., $g^{2}(0)=1$),
the corresponding field is coherent in nature. Similarly, $D_{a}(1)<0$
corresponds to the antibunching. Finally, substituting Eqs. (\ref{eq:4})-(\ref{eq:6})
into Eq. (\ref{Da}), we obtained the analytic expression for $D_{a}\left(1\right)$,
which is given by 

\begin{equation}
\begin{array}{lcl}
D_{a}(1) & = & 2|f_{3}|^{2}|\alpha_{2}|^{2}|\alpha_{1}|^{2}+\left\{ (f_{1}^{*}f_{5}\right.\\
 & + & 4f_{2}^{*}f_{3}+2f_{1}^{*2}f_{2}f_{3})|\alpha_{2}|^{2}|\alpha_{1}|^{2}\alpha_{1}^{*}\\
 & + & \left.f_{1}^{*2}f_{3}^{2}|\alpha_{2}|^{2}\alpha_{1}^{*2}+{\rm c.c.}\right\} .
\end{array}\label{eq:26}
\end{equation}

To investigate the quantum statistical properties, we plot $D_{a}(1)$
with respect to the dimensionless interaction time which is depicted
in Fig. \ref{fig:ant} using a smooth (blue) line. The negative values
of the quantity, shown in the Fig. \ref{fig:ant}, demonstrates the
existence of lower-order antibunching and thus, the presence of a
nonclassical character in optomechanical system. Therefore, phonons
of the vibrating mirror modes are observed to be antibunched for a
small period of interaction time. 

\subsection{Higher-order antibunching}

In this section, we investigate the higher-order antibunching phenomena.
There have been a numerous proposals for detecting the higher-order
nonclassical photon statistics. For example, criteria for higher-order
photon statistics are given by Lee \cite{CT_Lee1}, Agarwal and Tara
\cite{GS_Agarwal}, and the generalized criterion of many photon antibunching
is proposed by Lee \cite{C T Lee}. In this article, we follow the
definition of Pathak and Garcia \cite{Patha_Garcia}, which can be
viewed as one of variants of various equivalent criteria, to investigate$(l-1)$th
order antibunching of phonon. Specifically, the criterion for higher-order
antibunching is \cite{Patha_Garcia}

\begin{equation}
\begin{array}{rcccl}
D(l-1) & = & \langle a^{\dagger l}a^{l}\rangle-\langle a^{\dagger}a\rangle^{l} & < & 0.\end{array}\label{Higher order antibunching criteria}
\end{equation}

Using the closed form of analytic solution in Eqs. (\ref{eq:4})-(\ref{eq:10-1})
in the criteria (\ref{Higher order antibunching criteria}), we obtained
the corresponding $c$- number equation which yields 

\begin{figure}[h]
\begin{centering}
\includegraphics[scale=0.4]{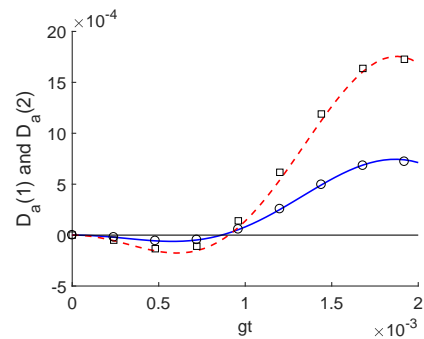}
\par\end{centering}
\caption{\label{fig:ant}(Color online) The presence of lower (for $l=2$)
and higher-order antibunching (for $l=3$) of phonons of vibrating
mirror mode is exhibited using solid (blue) and dashed (red) lines,
respectively. Here, we have amplified the lower-order quantity 5 times
to accommodate in the same plot. We have considered $|\alpha_{1}|^{2}=4$,
$|\alpha_{2}|^{2}=1$, and $\theta=\phi=0$.}
\end{figure}

\begin{widetext}

\begin{equation}
\begin{array}{lcl}
D(l-1) & = & (^{l}C_{2})^{2}|f_{2}|^{2}|\alpha_{1}|^{2l}+2l\,^{l}C_{2}|f_{2}|^{2}|\alpha_{1}|^{2(l+1)}+\left\{ \left(^{l}C_{2}f_{1}^{*}f_{5}+l\,^{l}C_{2}f_{2}^{*}f_{3}\right.\right.\\
 & + & \left.3\,^{l}C_{3}f_{1}^{*2}f_{2}f_{3}\right)|\alpha_{2}|^{2}|\alpha_{1}|^{2(l-1)}\alpha_{1}^{*}+\left(\frac{(2l-1)}{3}\,^{l}C_{2}f_{1}^{*}f_{4}+\frac{(3l-1)}{4}\,^{l}C_{3}f_{1}^{*2}f_{2}^{2}\right)|\alpha_{1}|^{2l}\\
 & + & \left.l\,^{l}C_{2}f_{1}^{*2}f_{2}^{2}|\alpha_{1}|^{2(l+1)}+^{l}C_{2}f_{1}^{*2}f_{3}^{2}|\alpha_{2}|^{2}|\alpha_{1}|^{2(l-2)}\alpha_{1}^{*2}+{\rm c.c.}\right\} ,
\end{array}\label{eq:27}
\end{equation}

\end{widetext}which is the analytic expression for the $(l-1)$th
order antibunching. The variation of the obtained result is depicted
in Fig. \ref{fig:ant} in dashed (red) line with corresponding lower-order
antibunching in smooth (blue) line. One can clearly observe that the
higher-order antibunching not only survives for relatively longer
period of time, also shows more depth of nonclassicality than corresponding
lower-order counterpart. This fact reestablishes the previous experimentally
reported result \cite{Maria-PRA-1,Maria-2,higher-order-PRL,with-Hamar}
that higher-order nonclassicality criteria may be useful in detecting
weaker nonclassicality.

As discussed in the previous section that results for optomechanics-like
system can be obtained by taking $\beta=0$ in Eqs. (\ref{eq:26})
and (\ref{eq:27}). Incidentally, in the present case, we failed to
observe antibunching in case of the\textcolor{red}{{} }trapped BEC system.
Therefore, here we have not included the graph that was obtained for
the BEC system.

\section{entanglement\label{sec6:entaglement}}

In this section, our main focus is to check whether the separable
initial state described by Eq. (\ref{eq:initial state}) evolves into
an entangled state due to interaction. To do so, we would investigate
the possibility of observing entanglement using a set of inseparability
criteria. In fact, there exist several inseparability criteria \cite{key-11}
(in terms of the annihilation and creation operators) that are suitable
to investigate the entanglement dynamics in this type of approach.
In this article, for simplicity, mostly we have used the Hillery-Zubairy
criteria \cite{key-4,key-6,key-7} to investigate the lower- and higher-order
entanglement and a criterion due to Duan et al. \cite{key-1} to investigate
only lower-order entanglement. For instance, the first criteria due
to Hillery and Zubairy \cite{key-4,key-6,key-7} is 

\begin{equation}
\begin{array}{lcl}
E_{1,a,k}^{1,1} & = & \langle a^{\dagger}(t)a(t)k^{\dagger}(t)k(t)\rangle-|\langle a(t)k^{\dagger}(t)\rangle|^{2}<0.\end{array}\label{eq:9}
\end{equation}
On the other hand, the second criteria due to Hillery and Zubairy
\cite{key-4,key-6,key-7} is

\begin{equation}
\begin{array}{lcl}
E_{2,a,k}^{1,1} & = & \langle a^{\dagger}(t)a(t)\rangle\langle k^{\dagger}(t)k(t)\rangle-|\langle a(t)k(t)\rangle|^{2}<0.\end{array}\label{eq:10}
\end{equation}
In the remaining part of this article we refer to these two criteria
as HZ-1 criterion and HZ-2 criterion, respectively. Additionally,
the inseparability criteria introduced by Duan et al. \cite{key-1}
is:

\begin{equation}
d_{a,k}=\langle(\Delta u)^{2}\rangle+\langle(\Delta v)^{2}\rangle-2<0,\label{eq:11}
\end{equation}
where 
\[
\begin{array}{lcl}
u & = & \frac{1}{\sqrt{2}}\{(a+a^{\dagger})+(k+k^{\dagger})\},\\
v & = & -\frac{i}{\sqrt{2}}\{(a-a^{\dagger})+(k-k^{\dagger})\}.
\end{array}
\]

The set of inseparability criteria listed in Eqs. (\ref{eq:9})-(\ref{eq:11})
are only sufficient, but not necessary. Thus, if any of these criteria
fails to detect the entanglement in a given state, it does not mean
that the state is separable, but obeying any of these criteria ensures
that the given state is entangled (inseparable). This establishes
our choice to use more than one such criteria to detect entanglement
in the quantum state.

We investigate the entanglement between movable mirror and the cavity
mode using HZ-1 criterion. Using Eqs. (\ref{eq:4})-(\ref{eq:10-1})
and (\ref{eq:9}), we obtain

\begin{widetext}

\begin{equation}
\begin{array}{lcl}
E_{1,a,k}^{1,1} & = & \langle N_{a}N_{k}\rangle-|\langle a(t)k^{\dagger}(t)\rangle|^{2}\\
 & = & |f_{3}|^{2}|\alpha_{2}|^{2}\left(3|\alpha_{2}|^{2}+1\right)+\left(|h_{2}|^{2}+|f_{2}|^{2}|\alpha_{1}|^{2}\right)|\alpha_{2}|^{2}|\alpha_{1}|^{2}\\
 & + & \left\{ f_{1}^{*}f_{3}|\alpha_{2}|^{2}\alpha_{1}^{*}+\left(f_{6}^{*}f_{1}+f_{1}^{*}f_{5}\right)|\alpha_{2}|^{2}|\alpha_{1}|^{2}\alpha_{1}^{*}+{\rm c.c.}\right\} .
\end{array}\label{eq:12}
\end{equation}

\end{widetext} The negative value of the right hand side of Eq. (\ref{eq:12})
is the witness of entanglement between the oscillating mirror and
the cavity modes which is shown in Fig. \ref{fig:The-entanglement}
a. The corresponding result for the BEC system is easily reducible
and is shown in Fig. \ref{fig:For-the-BEC} a. Specifically, we have
established possibility of observing entanglement between modes $a$
and $k$ in both the systems studied here. However, the presence of
entanglement depends on various parameters as shown in Figs. \ref{fig:The-entanglement}-\ref{fig:3D}.
Specifically, existence of entanglement between both the modes can
be controlled by controlling the phase of the oscillating mirror (cavity)
mode in optomechanical (optomechanics-like) system.

Similarly, using Eqs. (\ref{eq:4})-(\ref{eq:10-1}) and (\ref{eq:10}),
we obtained the analytic expression corresponding to HZ-2 criterion,
which is given by

\begin{widetext}
\begin{equation}
\begin{array}{lcl}
E_{2,a,k}^{1,1} & = & \langle N_{a}\rangle\langle N_{k}\rangle-|\langle a(t)k(t)\rangle|^{2}\\
 & = & |f_{3}|^{2}|\alpha_{2}|^{4}+\left(|\alpha_{1}|^{2}-1\right)|h_{2}|^{2}|\alpha_{2}|^{2}-\left\{ h_{1}h_{2}^{*}|\alpha_{2}|^{2}\alpha_{1}+f_{1}f_{4}^{*}|\alpha_{1}|^{4}|\alpha_{2}|^{2}\right.\\
 & + & \left.f_{1}f_{3}^{*}h_{1}^{*}h_{2}|\alpha_{2}|^{4}+\left(h_{1}h_{4}^{*}+f_{1}f_{2}^{*}h_{1}h_{2}^{*}\right)|\alpha_{2}|^{2}|\alpha_{1}|^{2}\alpha_{1}+{\rm c.c.}\right\} .
\end{array}\label{eq:32}
\end{equation}

\end{widetext} Variation of $E_{2,a,k}^{1,1}$ obtained through Eq.
(\ref{eq:32}) is illustrated in Fig. \ref{fig:The-entanglement}
b, and corresponding plot for the BEC system is shown in Fig. \ref{fig:For-the-BEC}
b. In analogy of the HZ-1 criterion, HZ-2 criterion also shows negative
values of the parameter $E_{2,a,k}^{1,1}$ (resulting in inseparability
of the two modes) depending up on the values of phase of the oscillating
mirror (cavity) mode in optomechanical (optomechanics-like) system.

Specifically, in Figs. \ref{fig:The-entanglement} a and \ref{fig:For-the-BEC}
a, we observed entanglement for $\theta=0$ and $\frac{\pi}{2}$,
but not for $\theta=\pi$, using HZ-1 criterion. On the other hand,
HZ-2 criterion of entanglement as shown in Figs. \ref{fig:The-entanglement}
b and \ref{fig:For-the-BEC} b could establish inseparability of two
modes for $\theta=\frac{\pi}{2},\,\pi$, while failed to detect for
$\theta=0$. Note that all these inseparability criteria are sufficient
in nature, not necessary, therefore, the negative values of the inequalities
conclude that the states are definitely entangled but fail to reach
to any conclusion for the positive values. 

To check the entanglement in the regions where HZ-1 and HZ-2 criteria
failed to detect them, we have used Duan et al.'s inseparability criteria.
Specifically, using Eqs. (\ref{eq:4})-(\ref{eq:10-1}) and (\ref{eq:11})
we obtained the analytic expression as

\begin{equation}
\begin{array}{lcl}
d_{a,k} & = & \left[|f_{2}|^{2}|\alpha_{1}|^{4}+\left(|f_{3}|^{2}+|h_{2}|^{2}\right)|\alpha_{2}|^{2}\right.\\
 & + & \left.\left\{ f_{2}h_{2}^{*}\alpha_{2}^{*}\alpha_{1}^{2}+{\rm c.c.}\right\} \right].
\end{array}\label{eq:33}
\end{equation}
The corresponding results for optomechanical and optomechanics-like
systems are shown in Fig. \ref{fig:The-entanglement} c and Fig. \ref{fig:For-the-BEC}
c, respectively. In this case, we failed to observe any nonclassicality
using this criteria. 

Apart from the phases of both the modes involved in interaction, one
can also control the possibility of entanglement generation by varying
the coupling constant $g$. This fact can be clearly established from
Fig. \ref{fig:3D} a and b. Specifically, we can see that the entanglement
detected using HZ-1 criterion (in Fig. \ref{fig:3D} a) tends to go
deeper with higher values of the coupling constant. In contrast, while
we use HZ-2 criterion we observed opposite nature (see Fig. \ref{fig:3D}
b) as the plot tends to become more positive. 

\begin{widetext}

\begin{figure}[h]
\begin{centering}
\begin{tabular}{cc}
\includegraphics[scale=0.4]{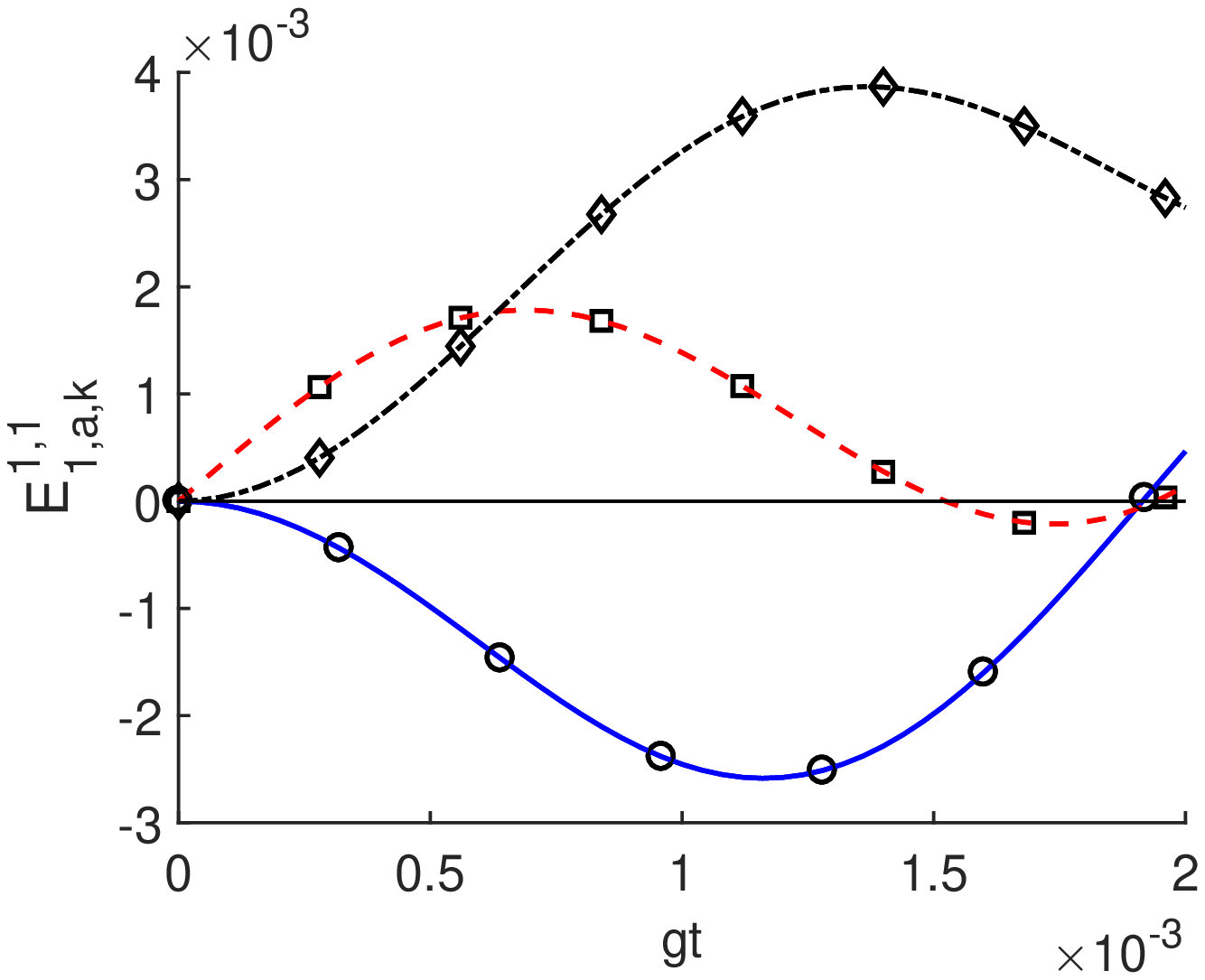} & \includegraphics[scale=0.4]{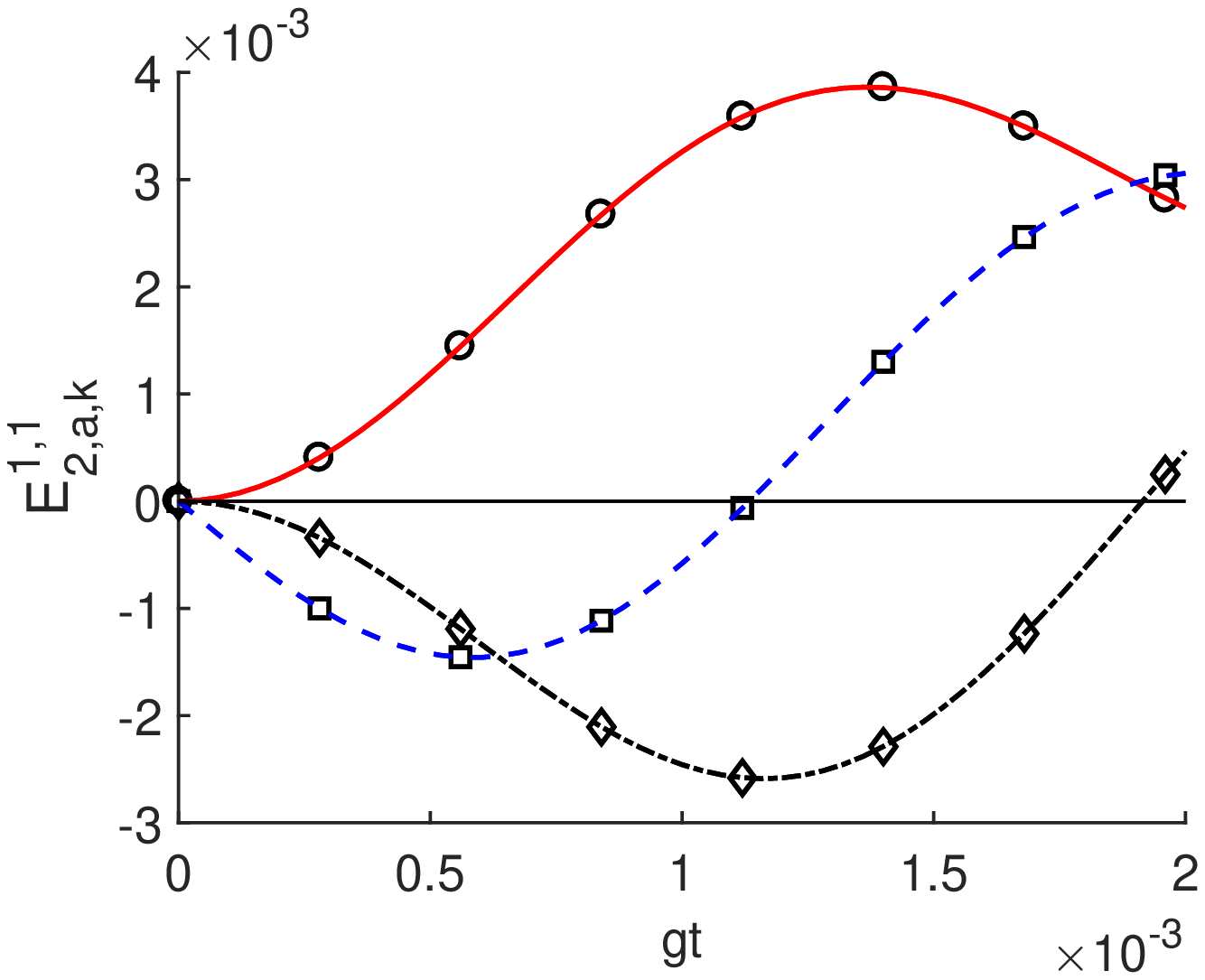}\tabularnewline
(a) & (b)\tabularnewline
\includegraphics[scale=0.4]{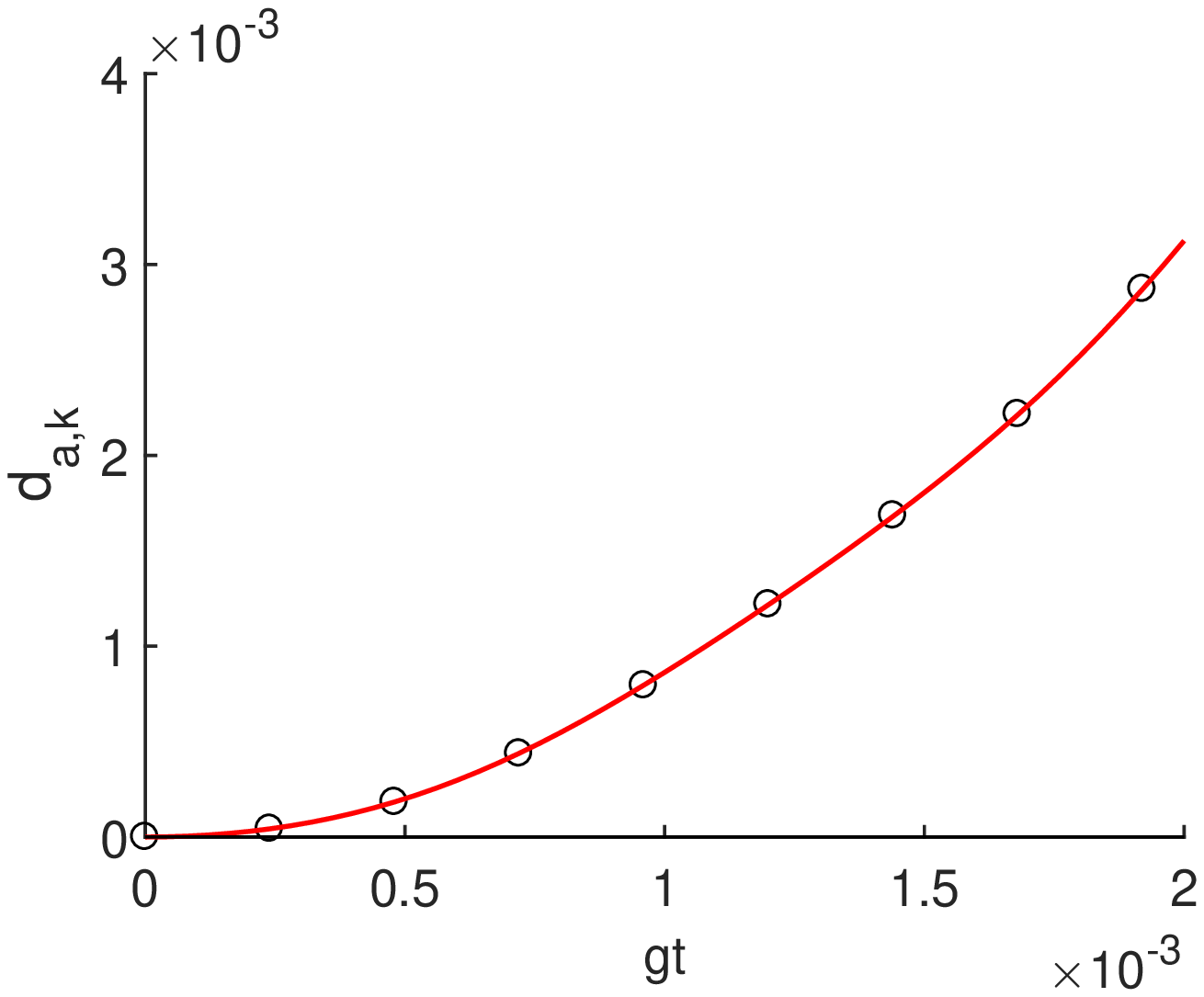} & \includegraphics[scale=0.4]{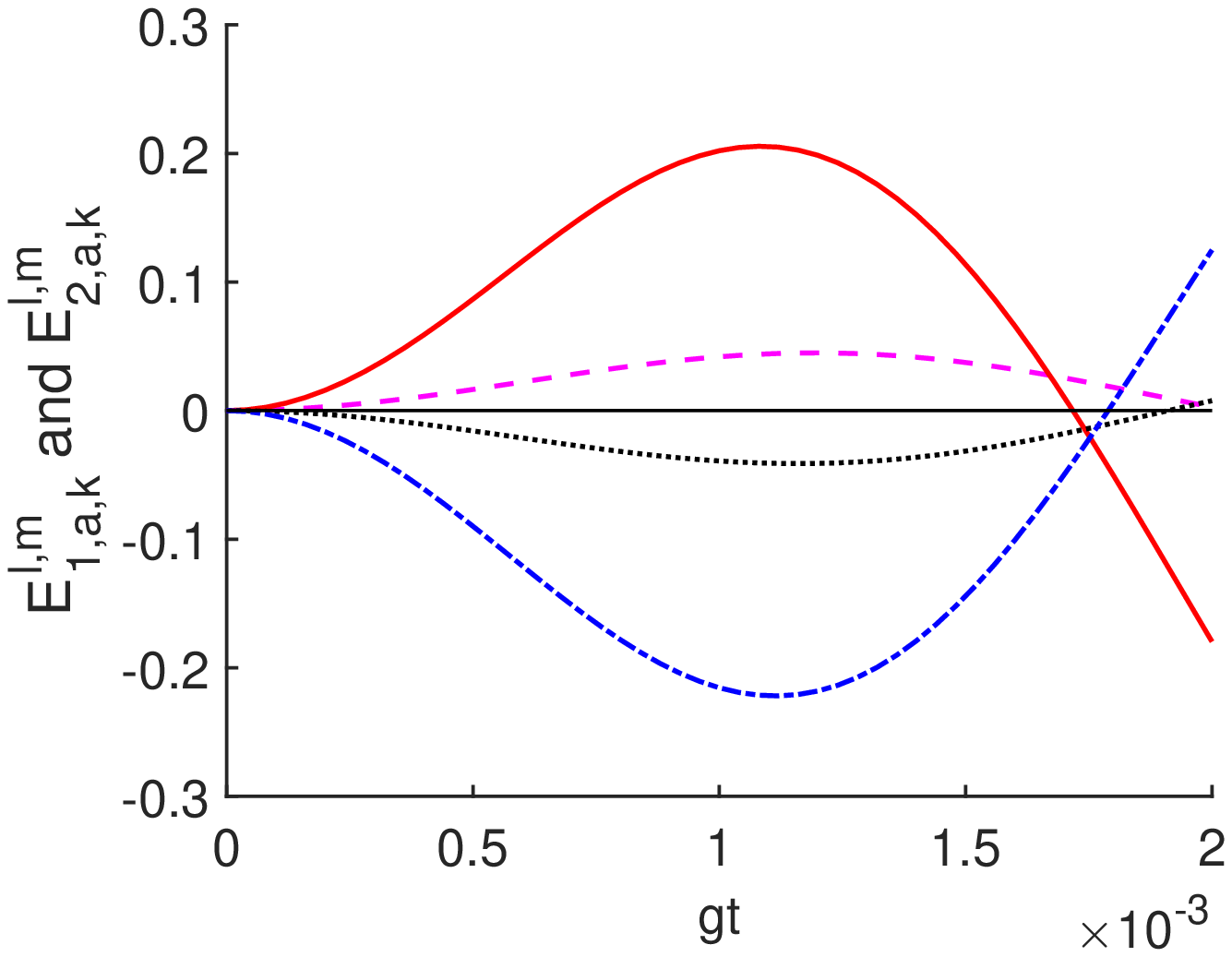}\tabularnewline
(c) & (d)\tabularnewline
\end{tabular}
\par\end{centering}
\caption{\label{fig:The-entanglement}(Color online) The intermodal entanglement
between cavity and mirror modes is illustrated for optomechanical
system with $|\alpha_{1}|^{2}=4$ and $|\alpha_{2}|^{2}=1$. In all
these plots, negative values of the shown quantities establish entanglement.
(a) The smooth (blue), dashed (red) and dash-dotted (black) lines
correspond to the variation of quantity for $\theta=0,\,\frac{\pi}{2}$
and $\pi$, respectively, using HZ-1 criterion. (b) Using HZ-2 criterion,
the smooth (red), dashed (blue), and dash-dotted (black) lines correspond
to $\theta=0,\,\frac{\pi}{2}$ and $\pi$, respectively. (c) Illustration
of the variation of Duan et al.'s entanglement parameter $d_{a,k}$.
(d) Higher-order entanglement is illustrated using HZ-1 and HZ-2 criteria
$\theta=\phi=0$, where dotted (black) line $(m=2,\,l=2)$, dashed-dotted
(blue) line $(m=2,\,l=3)$ correspond to HZ-1 criterion and smooth
(red) line $(m=2,\,l=3)$, dashed (magenta) line $(m=2,\,l=2)$ correspond
to HZ-2 criterion, respectively. }
\end{figure}



\begin{figure}[h]
\begin{centering}
\begin{tabular}{cc}
\includegraphics[scale=0.4]{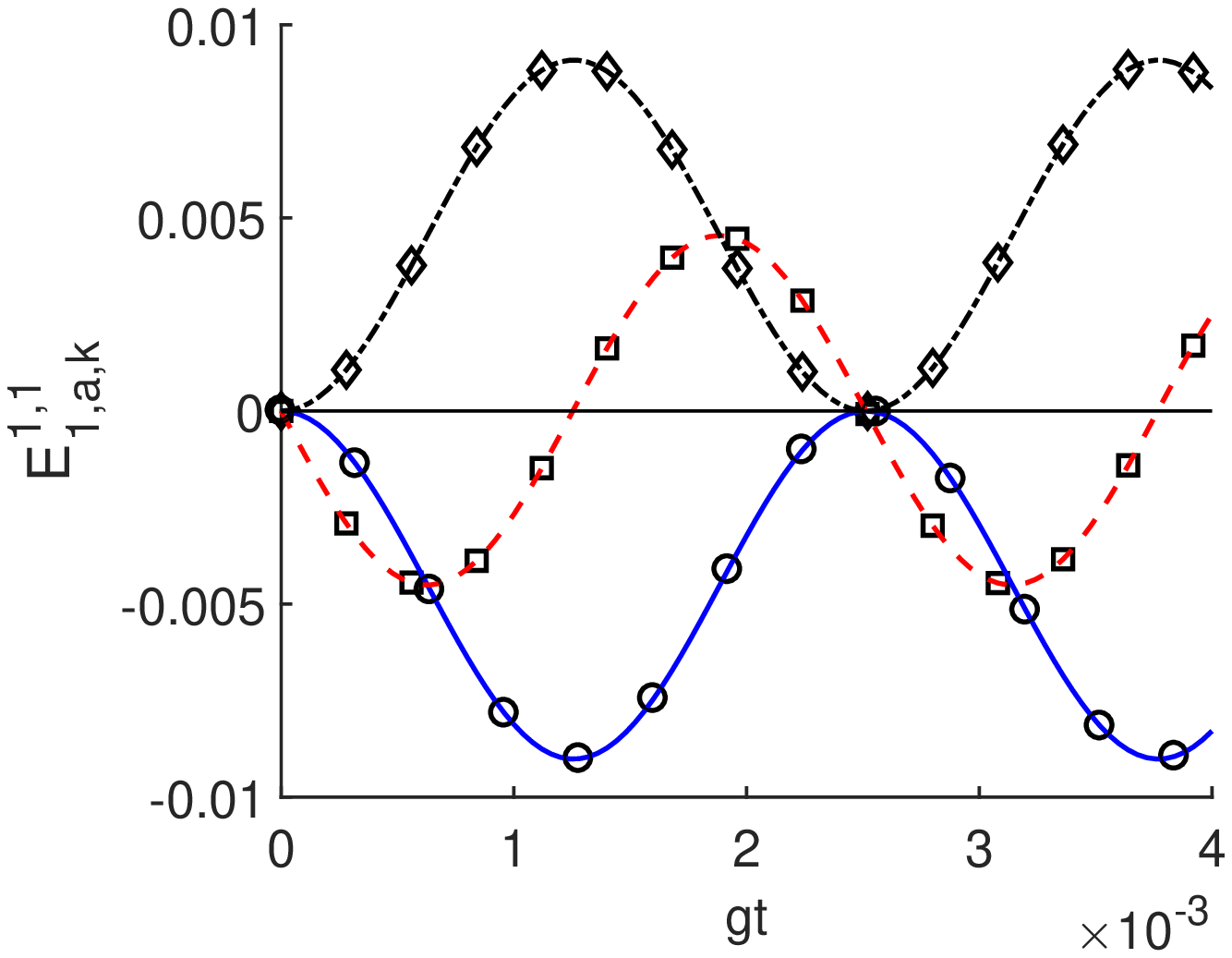} & \includegraphics[scale=0.4]{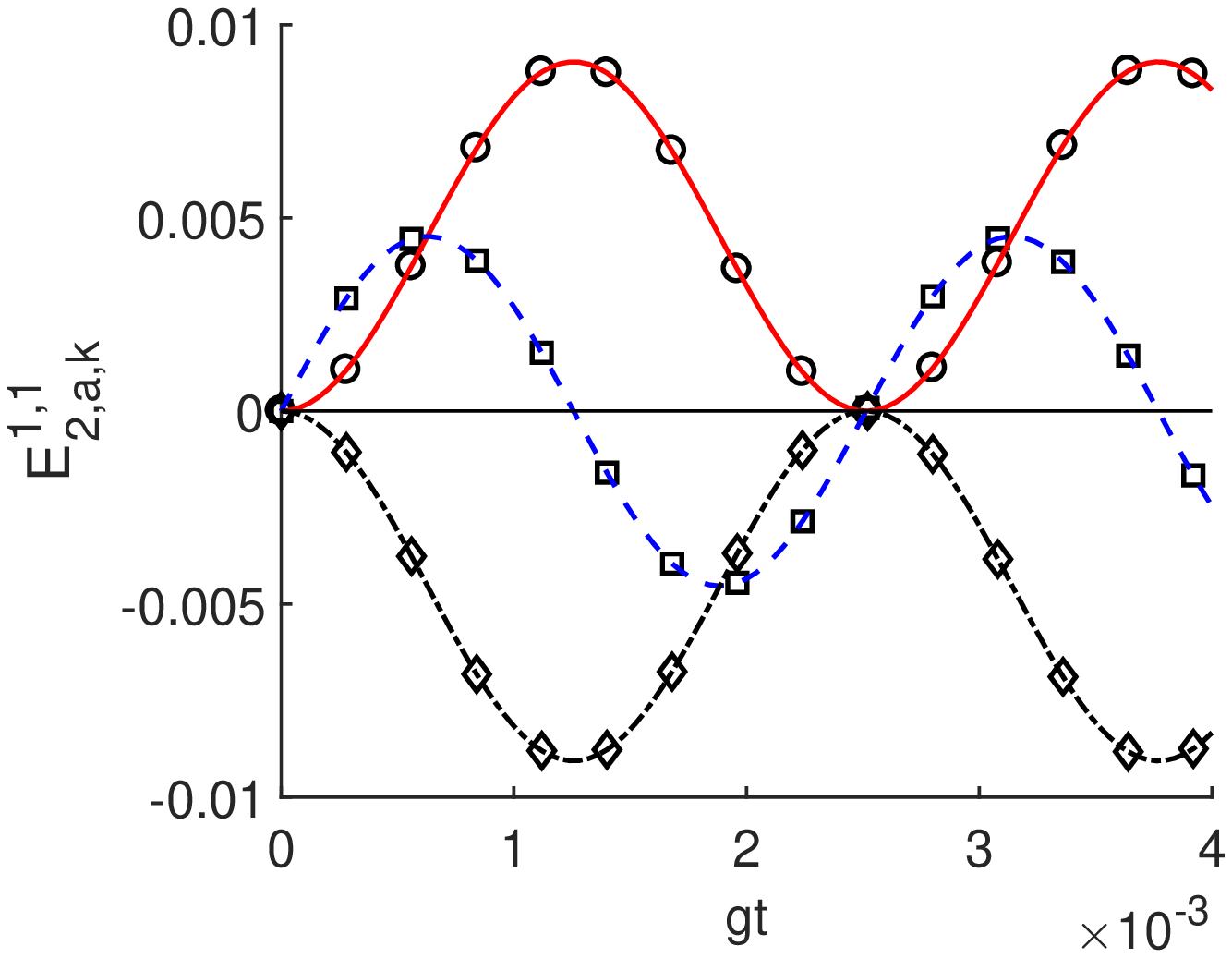}\tabularnewline
(a) & (b)\tabularnewline
\includegraphics[scale=0.4]{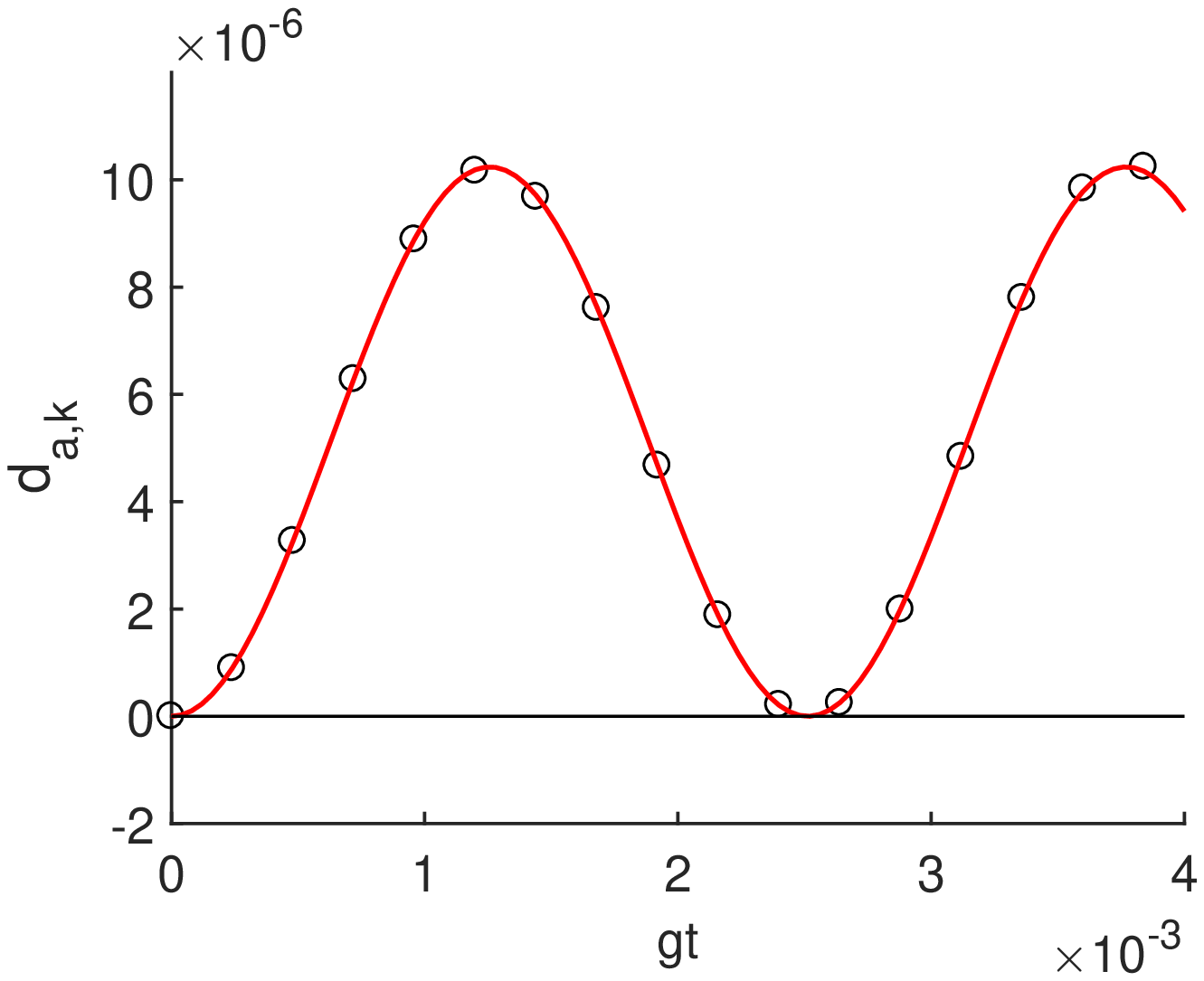} & \includegraphics[scale=0.4]{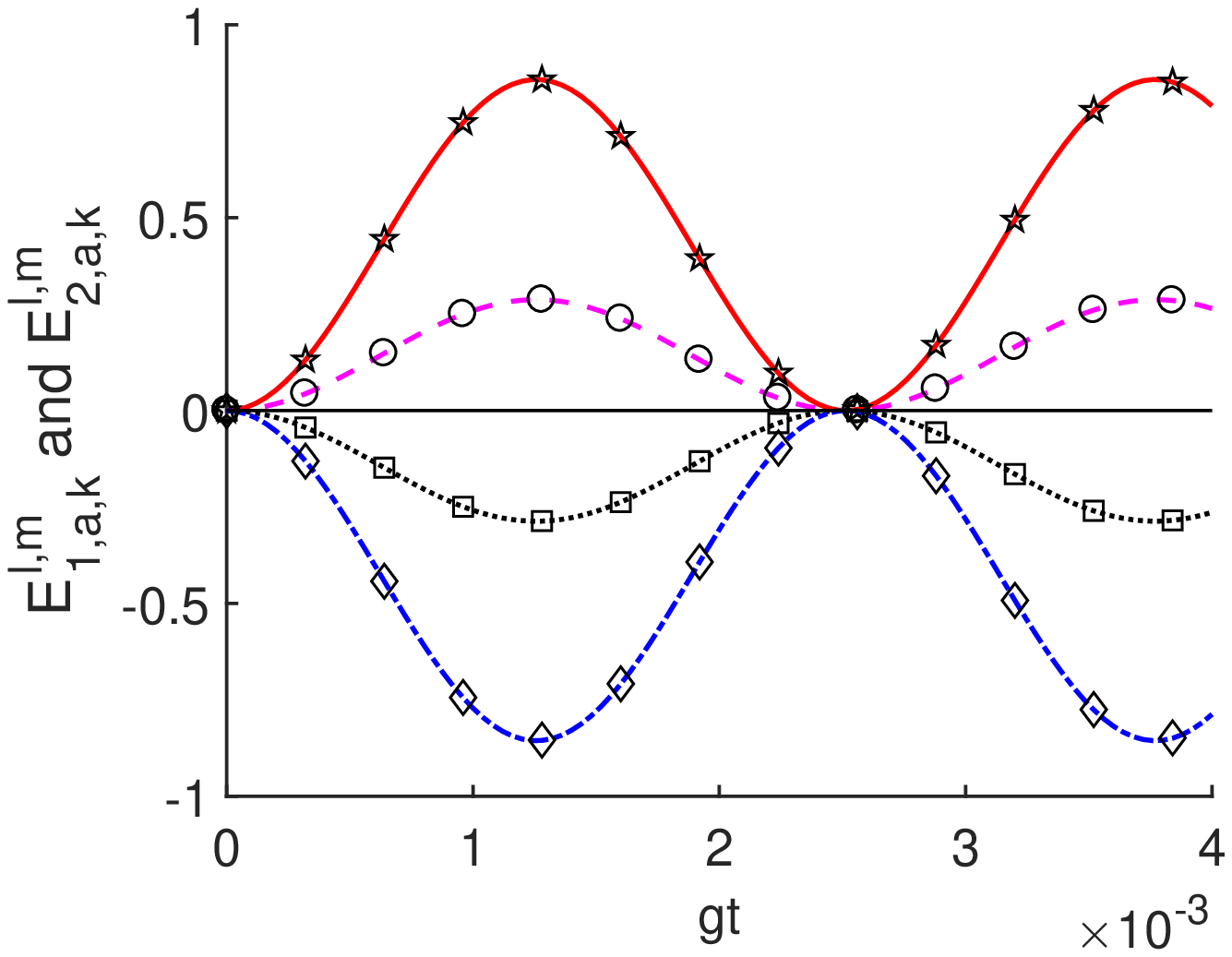}\tabularnewline
(c) & (d)\tabularnewline
\end{tabular}
\par\end{centering}
\caption{\label{fig:For-the-BEC}(Color online) For the BEC trapped optomechanics-like
system with excitation number $|\alpha_{1}|^{2}=2,$ $|\alpha_{2}|^{2}=4$,
(a) entanglement using HZ-1 criterion, where the smooth (blue), dashed
(red) and dash-dotted (black) lines correspond to $\theta=0,\,\frac{\pi}{2}$
and $\pi$, respectively. Similarly, (b) entanglement using HZ-2 criterion
with the smooth (red), dashed (blue) and dash-dotted (black) lines
corresponding to $\theta=0,\,\frac{\pi}{2}$ and $\pi$, respectively.
(c) Illustration of the variation of Duan et al.'s entanglement parameter
$d_{a,k}$. (d) Higher-order entanglement is illustrated with smooth
(red), dashed (blue) lines corresponding to $(m=2,\,l=2)$, $(m=2,l=3)$
using HZ-1 criterion, and dashed-dotted (blue), dotted (black) corresponding
to $(m=2,\,l=2)$, $(m=2,l=3)$ using HZ-2 criterion, respectively. }
\end{figure}



\begin{figure}[t]
\begin{centering}
\begin{tabular}{cc}
\includegraphics[scale=0.6]{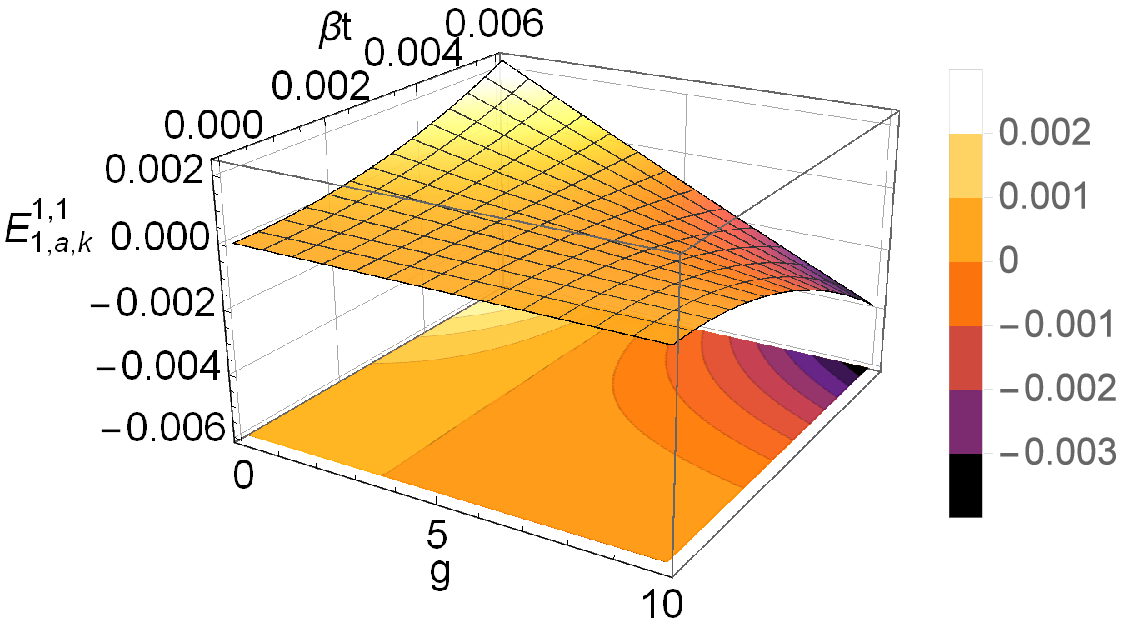} & \includegraphics[scale=0.6]{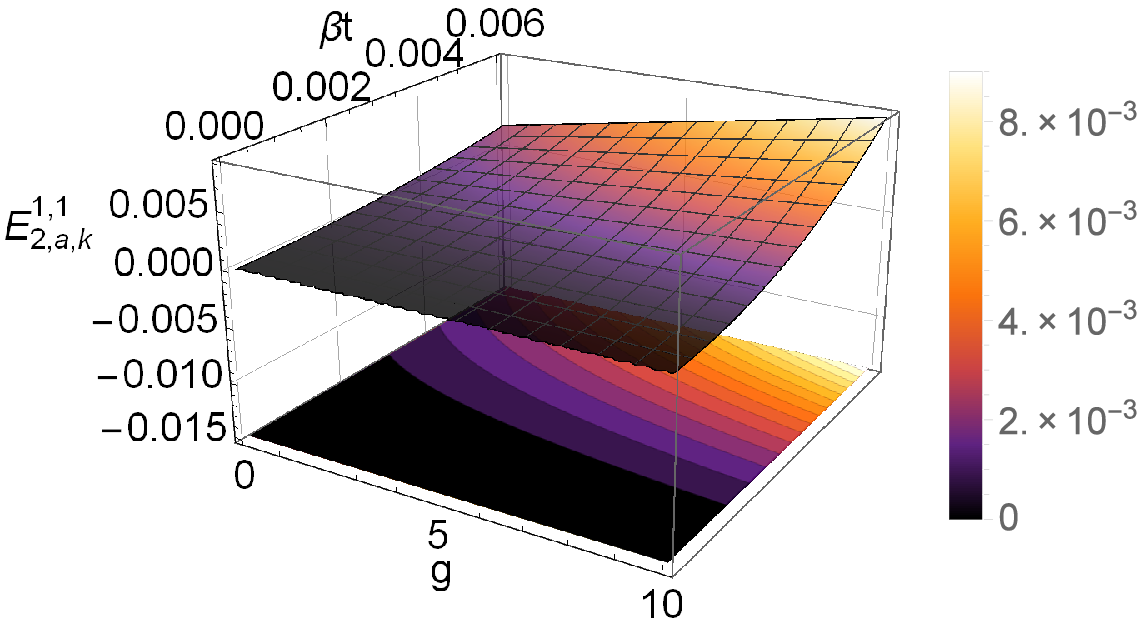}\tabularnewline
(a) & (b)\tabularnewline
\includegraphics[scale=0.6]{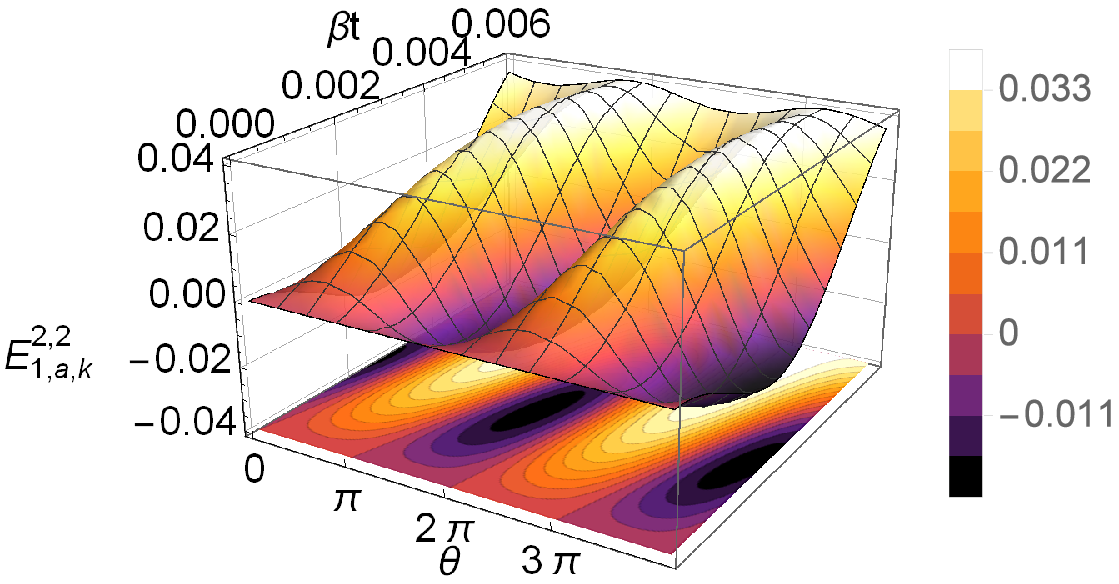} & \includegraphics[scale=0.6]{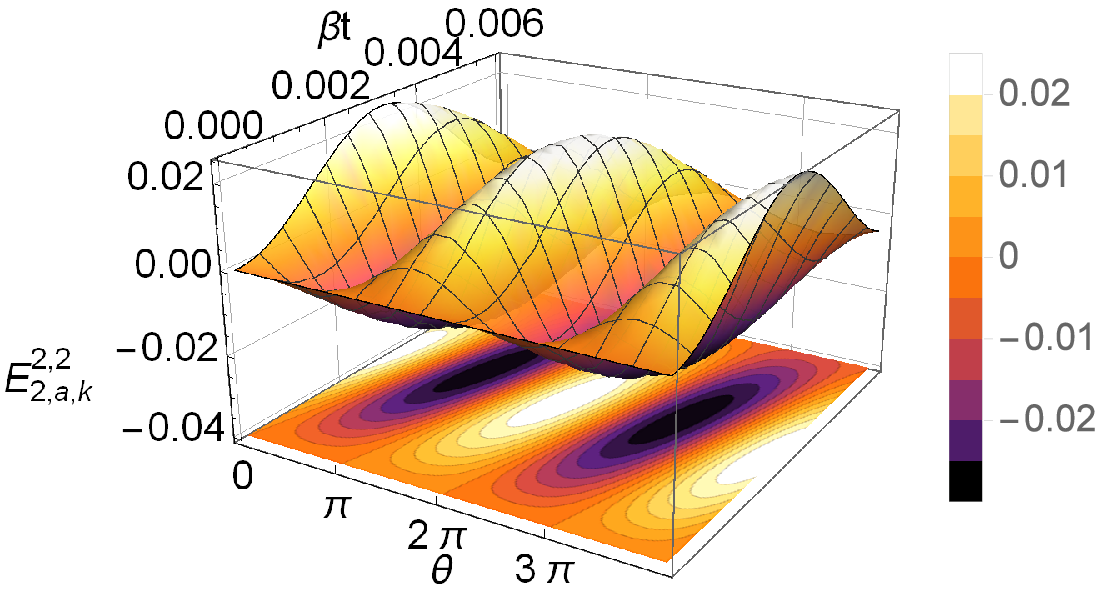}\tabularnewline
(c) & (d)\tabularnewline
\end{tabular}
\par\end{centering}
\centering{}\caption{\label{fig:3D}(Color online) Variation of entanglement for $|\alpha_{1}|=4$,
$|\alpha_{2}|=1$ with respect to interaction constant $g$ and rescaled
time using (a) HZ-1 and (b) HZ-2 criteria, respectively. Higher-order
entanglement varying with respect to the phase of the oscillating
mirror using (c) HZ-1 and (b) HZ-2 criteria, respectively. (d) using
HZ-2 criterion. Here, $\theta=\phi=0$ for (a) and (b), and $\phi=0$
and $g=2$ in (c) and (d).}
\end{figure}

\end{widetext}

\subsection{Higher-order entanglement}

Higher-order entanglement criteria ensure the possibility of detecting
the presence of weak nonclassical effects (entanglement) present in
the system. Therefore, here we use two criteria to investigate the
higher-order two mode entanglement introduced Hillery and Zubairy
\cite{key-4,key-6}. According to these criteria, the higher-order
entanglement exists if either

\begin{equation}
\begin{array}{lcl}
E_{1,a,k}^{l,m} & = & \langle a^{\dagger l}(t)a^{l}(t)k^{\dagger m}(t)k^{m}(t)\rangle-|\langle a^{l}(t)k^{\dagger m}(t)\rangle|^{2}<0\end{array}\label{eq:17}
\end{equation}
or

\begin{equation}
\begin{array}{lcl}
E_{2,a,k}^{l,m} & = & \langle a^{\dagger l}(t)a^{l}(t)\rangle\langle k^{\dagger m}(t)k^{m}(t)\rangle-|\langle a^{l}(t)k^{m}(t)\rangle|^{2}<0\end{array}\label{eq:18}
\end{equation}
is satisfied, where $l$ and $m$ are non zero integers with $l\geq1$
and $m\geq1$. It is clear from Eqs. (\ref{eq:17}) and (\ref{eq:18}),
for the lowest possible values of these integers, the higher-order
entanglement criteria would reduce to the HZ-1 and HZ-2 given in Eqs.
(\ref{eq:9}) and (\ref{eq:10}), respectively. The higher-order two
mode entanglement is present between two modes if $l$ and $m$ satisfy
the relation $l+m\geq3$. Here, from Eq. (\ref{eq:17}) and using
Eqs. (\ref{eq:4})-(\ref{eq:10-1}), we obtain the generalized higher-order
entanglement for HZ-1 criterion as

\begin{widetext}

\begin{equation}
\begin{array}{lcl}
E_{1,a,k}^{l,m} & = & \left\{ (2m+1)|\alpha_{2}|^{2}+m^{2}\right\} l^{2}|f_{3}|^{2}|\alpha_{2}|^{2m}|\alpha_{1}|^{2(l-1)}+m^{2}|h_{2}|^{2}|\alpha_{2}|^{2m}|\alpha_{1}|^{2l}\\
 & + & l^{2}|f_{2}|^{2}|\alpha_{2}|^{2m}|\alpha_{1}|^{2(l+1)}+\left[\left(mlf_{1}f_{3}^{*}-2m\,^{l}C_{2}h_{1}^{*}h_{5}\right)|\alpha_{2}|^{2m}|\alpha_{1}|^{2(l-1)}\alpha_{1}\right.\\
 & + & ml\left(f_{1}^{*}f_{6}+f_{1}f_{5}^{*}\right)|\alpha_{2}|^{2m}|\alpha_{1}|^{2l}\alpha_{1}+2\,^{l}C_{2}C_{2}^{m}f_{1}^{2}f_{3}^{*2}|\alpha_{2}|^{2m}|\alpha_{1}|^{2(l-2)}\alpha_{1}^{2}\\
 & - & \left\{ 2\,^{l}C_{2}(mh_{1}^{*}h_{7}+^{m}C_{2}h_{1}^{*2}h_{3}^{2}+^{l}C_{2}f_{1}^{2}f_{3}^{*2})+m^{2}l\left(l-1\right)f_{1}f_{3}^{*}h_{1}^{*}h_{3}\right\} |\alpha_{2}|^{2m}|\alpha_{1}|^{2(l-2)}\alpha_{1}^{2}\\
 & - & \left.ml(l-1)f_{1}f_{3}^{*}h_{1}^{*}h_{3}|\alpha_{2}|^{2(m+1)}|\alpha_{1}|^{2(l-2)}\alpha_{1}^{2}+{\rm c.c.}\right].
\end{array}\label{eq:36}
\end{equation}
In the similar manner, using Eqs. (\ref{eq:4})-(\ref{eq:10-1}) and
Eq. (\ref{eq:18}), we obtain the expression for the generalized higher-order
entanglement for HZ-2 criterion

\begin{equation}
\begin{array}{lcl}
E_{2,a,k}^{l,m} & = & m^{2}|h_{2}|^{2}|\alpha_{2}|^{2m}|\alpha_{1}|^{2l}+l^{2}|f_{3}|^{2}|\alpha_{2}|^{2(m+1)}|\alpha_{1}|^{2(l-1)}\\
 & - & \left\{ l\,^{l}C_{2}|f_{2}|^{2}|\alpha_{2}|^{2m}|\alpha_{1}|^{2l}\left(1+|\alpha_{1}|^{2}\right)+m^{2}l^{2}|h_{2}|^{2}|\alpha_{2}|^{2m}|\alpha_{1}|^{2(l-1)}\right\} \\
 & - & \left[\left(mlh_{1}h_{2}^{*}+2m\,^{l}C_{2}h_{1}h_{4}^{*}\right)|\alpha_{2}|^{2m}|\alpha_{1}|^{2(l-1)}\alpha_{1}\right.+ml\left(h_{1}h_{4}^{*}+f_{1}f_{2}^{*}h_{1}h_{2}^{*}\right)|\alpha_{2}|^{2m}|\alpha_{1}|^{2l}\alpha_{1}\\
 & + & 2^{l}C_{2}\left(mh_{1}h_{6}^{*}+^{m}C_{2}h_{1}^{2}h_{2}^{*2}\right)|\alpha_{2}|^{2m}|\alpha_{1}|^{2(l-2)}\alpha_{1}^{2}+\left(l^{2}f_{1}f_{4}^{*}+l\,^{l}C_{2}f_{1}^{2}f_{2}^{*2}\right)|\alpha_{2}|^{2m}|\alpha_{1}|^{2(l+1)}\\
 & + & \left.ml(l-1)f_{1}f_{3}^{*}h_{1}h_{2}^{*}|\alpha_{2}|^{2(m+1)}|\alpha_{1}|^{2(l-2)}\alpha_{1}^{2}+ml^{2}f_{1}f_{3}^{*}h_{1}^{*}h_{2}|\alpha_{2}|^{2(m+1)}|\alpha_{1}|^{2(l-1)}+{\rm c.c.}\right].
\end{array}\label{eq:37}
\end{equation}

\end{widetext}

It is a tedious job to obtain the compact analytic expressions reported
in Eqs. (\ref{eq:36})-(\ref{eq:37}), and also quite difficult to
interpret the results directly. However, it allows us to establish
the dependence of these quantities on various physical parameters.
Therefore, we plot the analytic results with various parameters as
shown in Fig. \ref{fig:The-entanglement} d and Fig. \ref{fig:For-the-BEC}
d for the optomechanical and BEC systems, respectively. Specifically,
in Fig. \ref{fig:The-entanglement} d, we have shown both HZ-1 and
HZ-2 higher-order entanglement criteria together and higher-order
entanglement is evidently present for most of the values of rescaled
time (except for very high values of rescaled time). Along the same
line, the BEC system also exhibits higher-order entanglement for all
the values of rescaled time except $2.5\times10^{-3}$ as at least
one of the lines showing variation of higher-order entanglement is
negative (cf. Fig. \ref{fig:For-the-BEC} d). 

Dependence of higher-order entanglement on various physical parameters
can also be established. Here, we show the variation of the HZ-1 and
HZ-2 higher-order entanglement parameter with the phase of the oscillating
mirror (cavity) mode in optomechanical (optomechanics-like) system
in Fig. \ref{fig:3D} c and d, respectively. The plots illustrate
that the possibility of detecting entanglement can be controlled by
the value of the phase parameter. It is important to note here that
the three dimensional variation of both lower- and higher-order entanglement
parameters with the coupling constant and phase parameter is similar.
Therefore, we are avoiding repetition and would like to emphasize
this point only in the text.

\begin{figure}[h]
\begin{centering}
\begin{tabular}{c}
\includegraphics[scale=0.4]{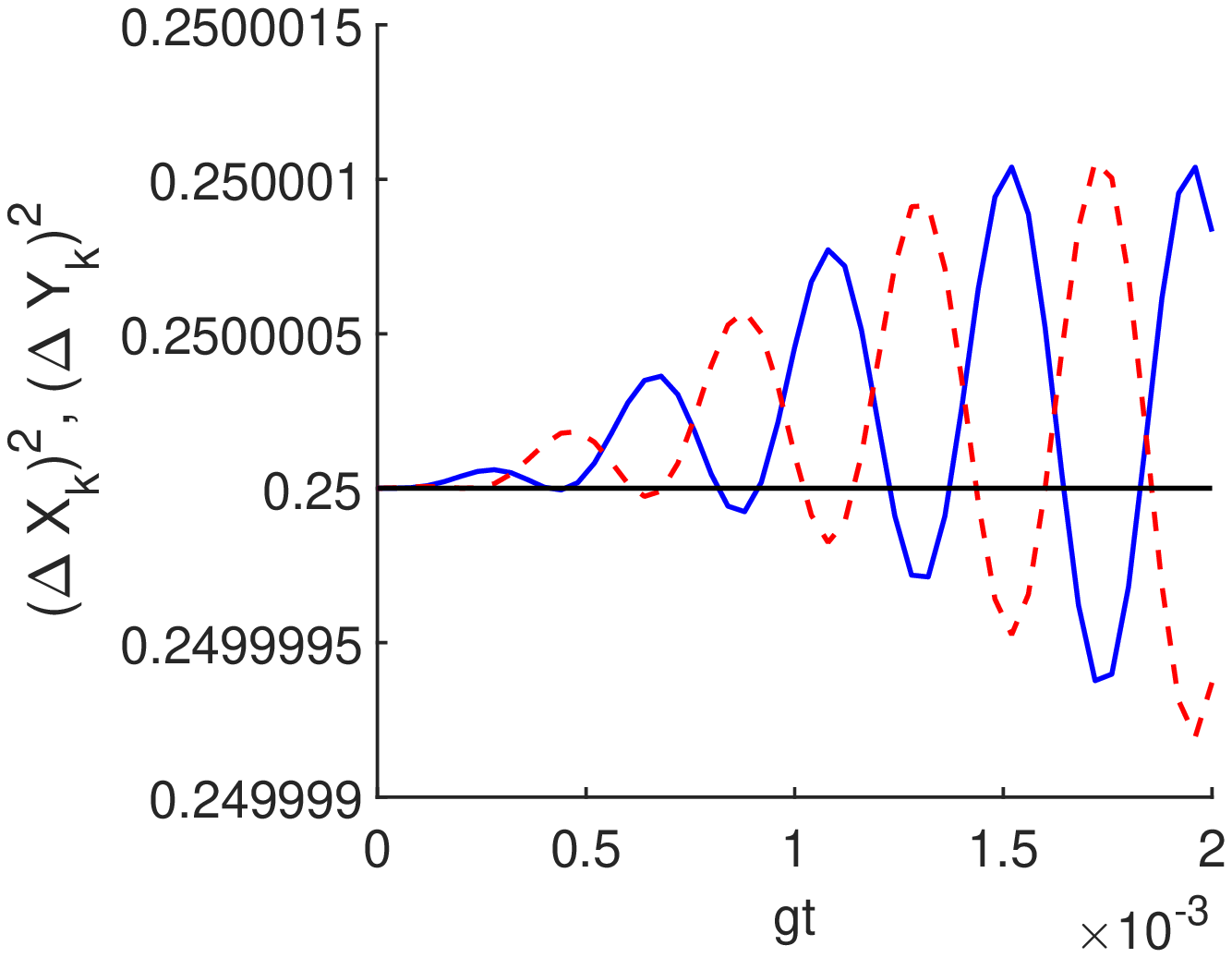}\tabularnewline
(a)\tabularnewline \\ \includegraphics[scale=0.4]{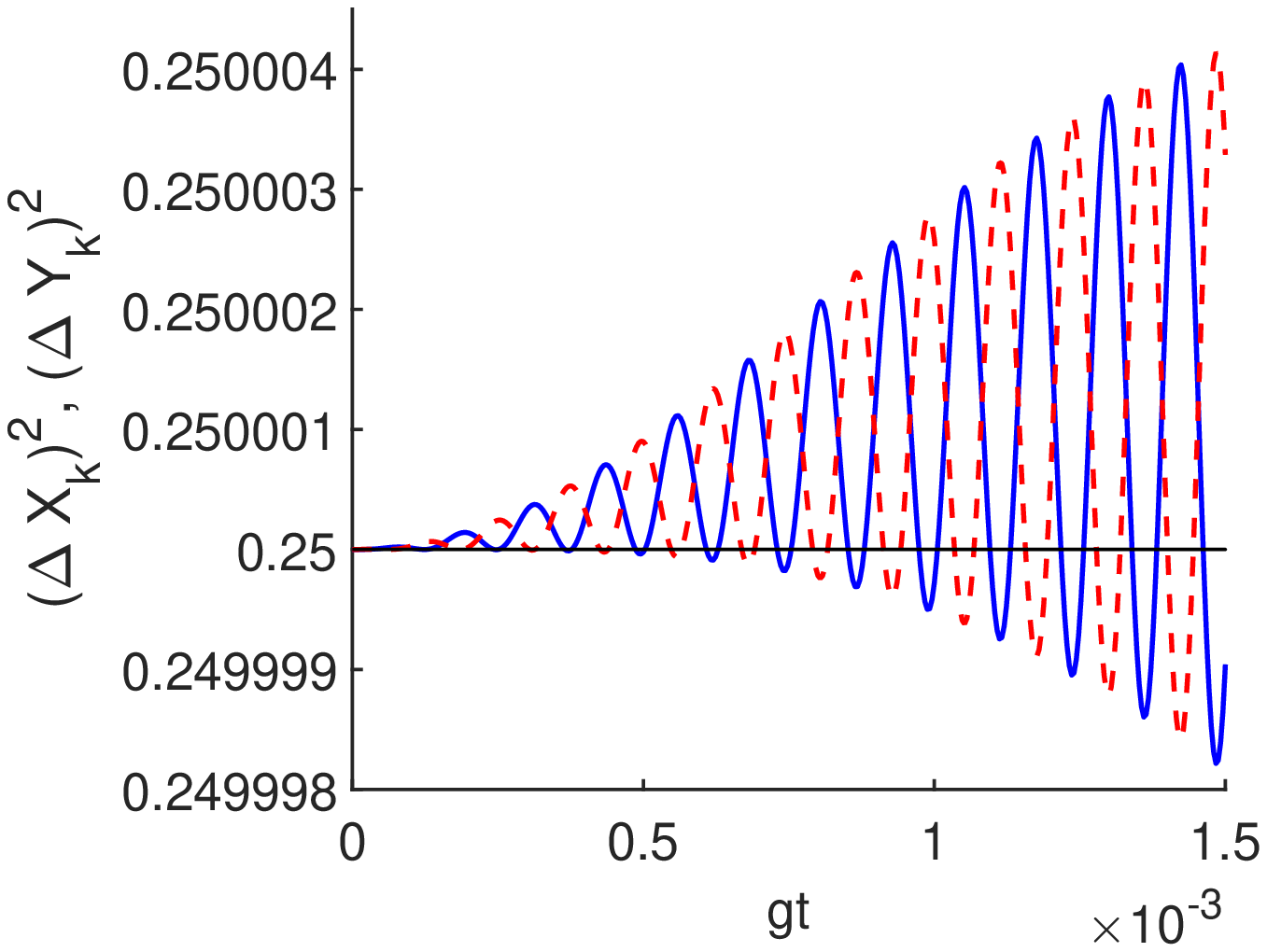}\tabularnewline
(b)\tabularnewline
\end{tabular}
\par\end{centering}
\caption{\label{fig:num}(Color online) The presence of quadrature squeezing
in mode $k$ for the (a) optomechanical and (b) BEC systems. In both
figures, the solid (blue) and dashed (red) lines correspond to the
variances in quadratures $X_{k}$ and $Y_{k}$, respectively. }
\end{figure}

Before we conclude the paper, it becomes imperative to discuss the
single mode nonclassicality present in the systems of our interest
that are not discussed so far. Specifically, we have not reported
squeezing in the cavity (trapped BEC) mode in the optomechanical (optomechanics-like)
system due to weak nonclassicality observed in them. Therefore, for
the sake of completeness of the present work, we are reporting here
only numerical results obtained in that case in Fig. \ref{fig:num}.
In brief, both the systems the corresponding modes show squeezing
phenomena only after an appreciable time evolution. On top of that
these modes fail to show antibunching. However, the nonclassical behavior
of the concerned mode has already been established with the help of
intermodal nonclassicality.

\section{conclusion\label{sec7:conclusion}}

In conclusion, we would like to stress on the fact that the work presented
here offers prospects for the observation of lower- and higher-order
nonclassical features present in the cavity with a nonlinearly movable
mirror comprising of an optomechanical system and in a BEC trapped
optomechanics-like system. The optomechanical system with nonlinearly
movable mirror can be considered as a Kerr-like nonlinear medium.
In the absence of nonlinearity, the system Hamiltonian reduces to
that of the BEC trapped optomechanics-like system. We first obtain
the solution for the generalized nonlinear system, subsequently, the
solution for the BEC system is obtained as the limiting case.

In the present study, we have assumed that there is no leakage of
photon through the cavity. Therefore, the main source of decoherence
is due to the interaction of the movable mirror with environment which
can be neglected up to some extent unless the mirror is heavily damped.
Moreover, for simplicity, we have assumed that the nonlinearity present
in the system is quartic in nature. The model Hamiltonian of the physical
system is constructed using the rotating wave approximation to eliminate
the non conserving energy terms. In Heisenberg picture, we obtain
the equation of motion of the corresponding field operators. Subsequently,
we obtain a perturvative analytic solution using Sen-Mandal technique.
Finally, we used this perturbative analytic solution to obtain analytic
expressions for various nonclassicality parameters and plot those
parameters obtain signatures of nonclassicality. Various types of
lower- and higher-order nonclassicality have been observed and they
are summarized in Table \ref{tab:Observation-of-nonclassicality}.
The validity of the obtained perturbative expressions for various
nonclassical parameters in the domain of study of time evolution is
also verified by comparing the obtained results with those obtained
as numerical solution of time dependent Schr{\"o}dinger equation.
Analytic and numerical results are found to be in good agreement.

As summarized in in Table \ref{tab:Observation-of-nonclassicality},
in this paper, we have used the obtained perturbative solution to
observe the possibility of generation of squeezed, antibunched and
entangled states in both the systems of our interest. Specifically,
both lower- and higher-order squeezing are observed in the movable
mirror mode in the optomechanical system. In contrast, the reduced
results for the optomechanics-like system could generate neither lower-order
nor higher-order squeezing in the corresponding cavity mode. However,
both the systems are found to demonstrate intermodal squeezing. Similarly,
lower- and higher-order antibunching is observed to be present (absent)
in the movable mirror (cavity) mode in the optomechanical (optomechanics-like)
system. It's particularly noteworthy that so many higher-order nonclassical
phenomena have been observed in these two systems. This is important
because of the facts that higher order nonclassicality have yet been
observed only in a limited number of physical systems and it helps
to identify very week nonclassiclaity.

In case of lower and higher-order entanglement using Hillery and Zubairy's
set of criteria, both the systems are found to show possibility of
inseparable states, which can be easily controlled by changing the
phase parameter for the movable mirror (cavity) mode in the optomechanical
(optomechanics-like) system. Additionally, Duan et al.'s criteria
of lower-order entanglement failed to detect nonclassicality in any
system.

\begin{widetext}

\begin{table}[h]
\begin{tabular}{lcc}
\hline 
 & %
\begin{tabular}{c}
\textbf{Cavity with}\tabularnewline
\textbf{a movable mirror}\tabularnewline
\end{tabular} & \textbf{Cavity with BEC}\tabularnewline
\hline 
\hline 
Squeezing  & Observed for mode $a$ & Not observed\tabularnewline
Higher-order squeezing & Observed for mode $a$ & Not observed\tabularnewline
Intermodal squeezing  & Observed between modes $a$ and $k$ & Observed between modes $a$ and $k$\tabularnewline
Antibunching & Observed for mode $a$ & Not observed for mode $a$\tabularnewline
Higher-order antibunching & Observed for mode $a$ & Not observed for mode $a$\tabularnewline
Entanglement using HZ-1 & Observed for mode $a$ for $\theta=0,\,\pi/2$ & Observed for mode $a$ for $\theta=0,\,\pi/2$\tabularnewline
Entanglement using HZ-2 & Observed for mode $a$ for $\theta=\pi/2,\,\pi$ & Observed for mode $a$ for $\theta=\pi/2,\,\pi$\tabularnewline
Entanglement using Duan criterion & Not observed & Not observed\tabularnewline
Higher-order entanglement using HZ-1 & Observed for $\theta=0$ & Observed for $\theta=0$\tabularnewline
Higher-order entanglement using HZ-2 & Observed for $\theta=0$ & Not observed for $\theta=0$\tabularnewline
\hline 
\end{tabular}

\caption{\label{tab:Observation-of-nonclassicality} Observation of nonclassicality
in cavity with a movable mirror and BEC trapped in a cavity are summarized
here.}
\end{table}

\end{widetext}

In brief, the present study not only revealed nonclassical features
present in both optomechanical and optomechanics-like systems, it
also established that the nonclassical features observed here can
be controlled by controlling various parameters, such as the phase
of the movable mirror and cavity modes. Note that the single mode
quadrature squeezing in the cavity and trapped BEC modes was not observed
for the smaller values of the rescaled time, but intermodal squeezing
involving these modes has been observed for the same values of the
rescaled time. Further, the presence of entanglement and its controllable
behavior opens up new doors of possibilities in both optomechanical
and optomechanics-like systems. Various types of nonclassicality observed
here can be employed in different quantum information processing tasks.

We can conclude the paper with a hope that the rich variety of nonclassical
behavior observed in both the cavity systems we have studied here
will soon be experimentally verified by growing experimental facilities
and escalated interest in the field of optomechanical and optomechanics-like
systems.
\begin{acknowledgments}
AP and NA thank Department of Science and Technology (DST), India
for the support provided through the project number EMR/2015/000393.
KT acknowledges support from the Council of Scientific and Industrial
Research (CSIR), Government of India. 
\end{acknowledgments}

\end{document}